\documentclass[12pt,aps,showpacs]{revtex4}
\usepackage{epsfig}
\usepackage{bbm}
\usepackage{amssymb}
\usepackage{amsmath}

\newcommand{\nc}{\newcommand}
\nc{\be}{\begin{equation}}
\nc{\ee}{\end{equation}}
\nc{\bea}{\begin{eqnarray}}
\nc{\eea}{\end{eqnarray}}
\nc{\nn}{\nonumber}

\nc{\rme}{\textrm{e}}
\nc{\rmi}{\textrm{i}}

\def\Slash#1{#1\kern-0.55em\raise.05ex\hbox{/}}
\def\slash#1{#1\kern-0.5em\raise.05ex\hbox{{$\scriptstyle /$}}}

\begin{document}
\rightline{CERN-PH-TH/2006-115, HD-THEP-06-09, ITP-UU-06/08, SPIN-06/06}

\title{ Electroweak Phase Transition and Baryogenesis in the nMSSM}

\author{Stephan J.~Huber$^{(1)}$, Thomas Konstandin$^{(2)}$, Tomislav Prokopec$^{(3)}$,
    Michael G. Schmidt$^{(4)}$
       }
\email[]{Stephan.Huber@cern.ch}
\email[]{Konstand@kth.se}
\email[]{T.Prokopec@phys.uu.nl}
\email[]{M.G.Schmidt@thphys.uni-heidelberg.de}

\affiliation{$(1)\,$ Theory Division, CERN,
	CH-1211 Geneva 23, Switzerland}

\affiliation{$(2)\,$Department of Physics, Royal Institute of Technology (KTH), 
	AlbaNova University Center, 
	Roslagstullsbacken 11, 106 91 Stockholm, Sweden}

\affiliation{$(3)\,$
             Institute for Theoretical Physics (ITF) \& Spinoza Institute,
             Utrecht University, Leuvenlaan 4, Postbus 80.195,
             3508 TD Utrecht, The Netherlands}

\affiliation{$(4)\,$Institut f\"ur Theoretische Physik, Heidelberg
University,
             Philosophenweg 16, D-69120 Heidelberg, Germany
}

\date{\today}

\begin{abstract}
We analyze the nMSSM with CP violation in the singlet sector. We study
the static and dynamical properties of the electroweak phase
transition. We conclude that electroweak baryogenesis in this model is
generic in the sense that if the present limits on the mass spectrum
are applied, no severe additional tuning is required to obtain a
strong first-order phase transition and to generate a sufficient
baryon asymmetry.  For this we determine the shape of the nucleating
bubbles, including the profiles of CP-violating phases. The baryon
asymmetry is calculated using the advanced transport theory to first
and second order in gradient expansion presented recently.  Still,
first and second generation sfermions must be heavy to avoid large
electric dipole moments.
\end{abstract}


\maketitle

%
%
\newpage
\section{Introduction}

Lately, electroweak baryogenesis (EWBG)~\cite{Kuzmin:1985mm} has again
attracted more attention, not least because new
collider data will hopefully provide information about the relevant
(supersymmetric?) physical parameters in the next years.

Electroweak baryogenesis relies on a strong first-order electroweak
phase transition as the source of out-of-equilibrium effects. During a
first-order phase transition, bubbles of the low-temperature (broken)
phase nucleate and expand to fill all space.  An important aspect in
the determination of the baryon asymmetry is the impact of transport
phenomena~\cite{Cohen:1994ss}. Without transport, CP-violating
currents would only be generated near the bubble wall profile of the
Higgs vevs. Close to the wall, sphaleron transitions are already
strongly suppressed by the mass of the W-bosons, so that
B-violating processes are inefficient in producing the observed baryon
asymmetry (BAU).  Early works that incorporated transport effects into
the EWBG calculus were based on the WKB
approach~\cite{Joyce:1994fu}. In this framework, a CP-violating shift
in the dispersion relation induces a force of second order in the
gradient expansion in the Boltzmann equation and leads to CP-violating
fermion densities in the symmetric phase. Later on, this formalism was
applied to the MSSM, where CP violation results from mixing effects in
the chargino sector
\cite{JCK}.  In this context, the formalism had to be extended to the
case of mixing fermions.  
In the MSSM, the second-order source is too weak to
yield a successful baryogenesis~\cite{Kainulainen:2001cn,kpss}.

One disadvantage of the WKB method is the neglect of dynamical flavor
mixing effects. While the shift in the dispersion relation is due to
mixing of left- and right-handed components of the fermions and
already present in the one-flavor case, flavor mixing contributions
have been completely neglected after a flavor basis transformation to
the mass eigenbasis. A series of papers~\cite{cmqsw, cmqsw2} aimed at
improving on this point by including flavor mixing by using a
perturbative expansion of the Kadanoff--Baym equations. Here, the
deviations of the Green function have been interpreted as sources in
the diffusion equation. This approach --- like the WKB-approach --- has
the weakness that the transport equations only describe the dynamics
of two classical quasi-particles in the chargino sector. 
CP violation is communicated from the charginos to
the SM particles by their interactions.
Therefore, the authors of Refs.~\cite{cmqsw,
cmqsw2} used the Winos and Higgsinos as quasi-particles in the interaction
basis, where the interactions take a particularly simple form.  In the
WKB-approach the natural choice is the mass eigenbasis.  This dependence on a flavor
basis is unsatisfactory, especially since the flavor mixing
CP-violating source vanishes in the mass eigenbasis completely.   
A numerical analysis, making use of the first-order source in
the interaction basis in Refs.~\cite{cmqsw, cmqsw2}, leads to successful
electroweak baryogenesis for a certain range in the parameter space, 
even though at least some fine-tuning is required to fulfill the
electron electric dipole moment (EDM) constraints.

Recently, some of the authors have derived semiclassical transport
equations for the chargino sector from first
principles~\cite{Kainulainen:2001cn,kps}. The derivation is based on
the Kadanoff-Baym equations and does not depend on classical reasoning
to fill the gap between CP violation and transport
effects. Technically, the two main improvements on the resulting
transport equations are independence of the flavor basis and the
absence of the source strength ambiguities. First, since the
semiclassical transport equations describe the dynamics of a
$2\times2$ matrix in flavor space, the transformation properties of the
transport equations under flavor basis changes are explicit, and no
restriction to quasi-particles has to be used.  Secondly, because the
CP-violating sources appear naturally and uniquely as higher order
terms in the gradient expansion of the mass background fields, there
are no ambiguities in the source strength. The mixing effects lead to
an additional force of first order of the gradient expansion in the
transport equations.  In contrast, in the work reported on in \cite{cmqsw}, for
dimensional reasons, the sources had to be multiplied by a typical
thermalization time $\tau$, while in~\cite{cmqsw2} the removal of the
ambiguity in the source strength was based on the (classical) Fick's
law. In the case of second-order effects, the first principle derivation
confirms the WKB approach if the latter is handled
carefully~\cite{Joyce:1999fw,FH06}.

Applying this advanced transport theory to the
MSSM~\cite{kpss}, the quantitative analysis shows two distinct
features, which are less definitive in the results of the former
approach \cite{cmqsw, cmqsw2}: First, mixing effects are strongly 
suppressed away from mass
degeneracy in the chargino sector,
$|m^2_{\tilde\chi^\pm_1}-m^2_{\tilde\chi^\pm_2}| > (20 \textrm{
GeV})^2$. So mixing in that sector is only effective if the
{\it a priori} unrelated Wino mass parameter $M_2$ and the Higgsino
mass parameter $\mu$ are tied together. Secondly, the produced BAU
suffers from an exponential Boltzmann suppression in the case of heavy
charginos. In this formalism, successful electroweak baryogenesis
requires rather large CP violation in the chargino sector,
$\sin(\delta_{CP}) > 0.25$, even for the most favorable choice of the
other model parameters. In comparison, the approach followed in
Refs.~\cite{cmqsw,cmqsw2} leads to viable baryogenesis for less
constrained chargino masses and CP violation of order
$\sin(\delta_{CP}) > 0.1$. Hence, if the advanced transport
theory is used, not only the parameter space of viable baryogenesis is much
more restricted, but also the maximally achievable BAU is
smaller. Because of the necessity of large CP-violating phases, 
additional arguments (cancellations, or a large value of the CP-odd Higgs mass parameter)  
are required to suppress the electron EDM by a {\it factor 5-6}. On top of this, a light
right-handed stop and a light Higgs are needed to allow for a strong
first-order phase transition~\cite{Moreno:1998bq} (light stop
scenario).  Thus, electroweak baryogenesis in the MSSM is severely
constrained.

In this paper we study electroweak baryogenesis in a singlet extension of the MSSM,
where the divergences of the singlet tadpole are tamed by a discrete R-symmetry
\cite{PT98,PT99,Panagiotakopoulos:2000wp,DHMT00}. The R-symmetry is violated by
the supersymmetry breaking terms. A singlet tadpole is then induced at some high
loop order, which is too small to destabilize the weak scale, but large enough to evade
the cosmological domain wall problem \cite{ASW95}. 

Our analysis supports the result of Ref.~\cite{Menon:2004wv} that a strong
first-order phase transition is quite generic, once experimental constraints
on the Higgs and sparticle spectrum are taken into account. The phase transition
is induced by tree-level  terms in the Higgs potential without the need of a
light stop. Going beyond
Ref.~\cite{Menon:2004wv}, we actually compute the baryon asymmetry
and the bubble wall properties. We find that the observed baryon
asymmetry can be produced with mild tuning of the model parameters.
The first and second generation squarks and sleptons have to be heavy (a few TeV)
to suppress the one-loop contributions to the EDMs. In contrast to the MSSM,
there are no strong constraints from the two-loop EDMs, since $\tan{\beta}$
is usually small~\cite{Chang:2002ex,Pilaftsis:2002fe}. 
 
Several variants of MSSM singlet extensions have been studied in the literature
with respect to their impact on electroweak baryogenesis 
\cite{P92,DFM96,HS98,Huber:2000mg,BHKRV00,KLT04,Funakubo:2005pu}.
The most detailed of these studies is Ref.~\cite{Huber:2000mg}, where 
a general singlet model without discrete symmetries was considered.
This general model supports electroweak baryogenesis in a large part
of its parameter space. 
In the current work, the R-symmetry forbids a self coupling
of the singlet, leading to a quite constrained Higgs and neutralino phenomenology. 
Still, it is encouraging to see that even this restricted framework 
allows for successful electroweak baryogenesis. 

The paper is organized as follows. In section~\ref{model} we will 
present the model and clarify notation. In sections~\ref{ewbg} and \ref{ewpt},
we will discuss the mechanism that drives baryogenesis in the nMSSM and 
the dynamics of the phase transition. 
In section~\ref{numana}, numerical results will be presented before 
we conclude in section~\ref{concl}. 

\section{The Nearly Minimal Supersymmetric Standard Model\label{model}}
\subsection{The Model}

The notation in this section follows 
Refs.~\cite{Panagiotakopoulos:2000wp,Menon:2004wv}, including, however, a generalization 
to an additional CP-violating phase in the singlet sector.
The superpotential, including the multi-loop generated tadpole term is 
\bea
W_{\rm nMSSM} = \lambda \hat S \hat H_1 \cdot \hat H_2
- \frac{m^2_{12}}{\lambda} \hat S 
+ y_t \hat Q \cdot \hat H_2 \hat U^c + \cdots,
\label{W_nMSSM}
\eea
where the dots denote the remaining terms in the MSSM superpotential, 
$\hat H_1 = ( \hat H_1^0, \hat H_1^- )$, $\hat H_2 = ( \hat H_2^+, \hat H_2^0 )$,
$\hat S$ is the singlet superfield and, 
$A\cdot B = \epsilon^{ab} A_a B_b = A_1 B_2 - A_2 B_1$.

The tree-level potential consists of
\be 
V_0 = V_F + V_D + V_{\rm soft},
\ee 
where, restricting to third generation quarks, 
\bea
V_F &=& | \lambda H_1 \cdot H_2 - \frac{m^2_{12}}{\lambda}|^2
+ |\lambda S H_1^0 + y_t \tilde t_L \tilde t_R^* |^2 \nn \\
&& + |\lambda S H_1^- + y_t \tilde b_L \tilde t_R^* |^2
+ |\lambda S|^2 H_2^\dagger H_2 \nn \\
&& + |y_t \tilde t^*_R|^2 H_2^\dagger H_2 +|y_t \tilde Q \cdot H_2|^2, \nn \\
V_D &=& \frac{\bar g^2}{8} (H_2^\dagger H_2 - H_1^\dagger H_1)^2
+ \frac{g^2}{2} |H_1^\dagger H_2|^2, \\
V_{\rm soft} &=& m_1^2 H_1^\dagger H_1 + m_2^2 H_2^\dagger H_2 +
	m_s^2 |S|^2 \nn \\
	&&+ (t_s S + \, h.c.) + (a_\lambda S H_1 \cdot H_2 + \, h.c.) \nn \\
	&&+ m_Q^2 \tilde Q^\dagger \tilde Q + m_U^2 |\tilde t_R|^2
       	+ (a_t \tilde Q \cdot H_2 \tilde t_R^* + \, h.c.). \nn	 
\label{V_soft}
\eea
The Higgs sector of this
potential has only one physical CP-violating phase, which after some
redefinition of the fields can be attributed to the parameter
$t_s$. We assume that this phase is the only source of CP violation in
the model, i.e.~the gaugino masses and squark and slepton soft terms
are taken to be real as well as the parameter $\lambda$.

In the case when the squarks have vanishing vevs, the tree-level Higgs potential is 
\bea
V_0 &=& m_1^2 H_1^\dagger H_1 + m_2^2 H_2^\dagger H_2 +  m_s^2 |S|^2
	+ \lambda^2 |H_1 \cdot H_2 |^2 \nn \\
&& + \lambda^2 |S|^2 (H_1^\dagger H_1 + H_2^\dagger H_2)
+ \frac{\bar g^2}{8} (H_2^\dagger H_2 - H_1^\dagger H_1)^2 + 
\frac{g^2}{2} |H_1^\dagger H_2|^2 \nn \\
&& +\, t_s (S + \, h.c.) + (a_\lambda S H_1 \cdot H_2 + \, h.c.)
- m_{12}^2 (H_1 \cdot H_2 + \, h.c. ). 
\eea
We define the vevs as $\left<H_1^0\right> = \phi_1 e^{i\,q_1}$,
$\left<H_2^0\right> = \phi_2 e^{i\,q_2}$, $\left<S\right> = \phi_s
e^{i\,q_s}$. We use the gauge freedom to set
$\left<H_1^-\right> = 0$ and choose the phase convention
$q_1 = q_2 = q/2$. The absence of a charged condensate
$\left<H_2^+\right> = 0$ will be ensured by the positivity of the
squared charged Higgs mass \cite{Menon:2004wv} that is determined 
in the numerical analysis.

Using the definition of the $\beta$ angle 
\bea
 \phi_1 = \phi \, \cos(\beta),&& \quad \phi_2 = \phi \, \sin(\beta), 
\eea
and $t_s = |t_s| \, e^{i \, q_t}$, the tree-level potential reads finally 
\bea
V_0 = M^2 \phi^2 + m_s^2 \phi_s^2 
+ 2 \,  |t_s| \phi_s \, \cos( q_t + q_s ) + 2\, \tilde a  \phi^2 \phi_s
+ \lambda^2  \phi^2 \phi_s^2 + \tilde \lambda^2 \phi^4, \label{tree-pot}
\eea
where we have used
\bea
M^2 &=& m_1^2 \cos^2 \beta + m_2^2 \sin^2 \beta 
- m_{12}^2 \sin 2 \beta \cos q, \nn \\
\tilde a &=& \frac{a_\lambda}{2} \, \sin 2 \beta \cos(q+q_s),\nn \\
\tilde \lambda^2 &=& \frac{\lambda^2}{4} \sin^2 2\beta + 
\frac{\bar g^2}{8} \cos^2 2\beta,
\eea
to shorten the notation.

\subsection{Effective Potential at Zero Temperature}

At zero temperature we take into account, in addition to the tree-level potential, 
the Coleman--Weinberg one-loop contributions 
\bea
\Delta V = \frac{1}{16 \pi^2} \left[ \sum_b g_b \, h(m^2_b) 
- \sum_f g_f \, h(m^2_f)  \right], \label{C-W-log}
\eea
where the two sums run over the bosons and the fermions with the
degrees of freedom $g_b$ and $g_f$ respectively, and 
\bea
h(m^2) = \frac{m^4}{4} \left[ \ln \left( \frac{m^2}{Q^2} \right) - \frac32 \right].
\eea
We choose the renormalization point to be $Q=150$ GeV in the
$\overline{DR} $-scheme and suitable counter-terms, such that
the one-loop contributions to the potential preserve the location of the
tree-level minimum.  This leads  to the following
shifts in the bare mass parameters 
\bea
\Delta m^2_1 = - \frac{1}{2 \phi_1} \frac{\partial \Delta V}{\partial {\phi_1}},
\quad \Delta m^2_2 = - \frac{1}{2 \phi_2} \frac{\partial \Delta V}{\partial {\phi_2}},
\quad \Delta m^2_s = - \frac{1}{2 \phi_s} \frac{\partial \Delta V}{\partial {\phi_s}}.
\eea
As degrees of freedom of  the relevant one-loop contributions, we take 
\bea
g_W = 6, \quad g_Z = 3, \quad g_t = 12, \quad g_{\tilde t_L} = g_{\tilde t_R} = 6.
\eea
Contributions from the charginos and neutralinos are not taken
into account.  The masses used in the one-loop potential are listed in
Appendix~\ref{app_masses}.  The neutral Higgs masses are computed
from the second derivatives of the one-loop potential.

\subsection{Effective Potential at Finite Temperature\label{temp}}

Taking into account temperature effects, we correct the effective potential 
by the thermal one-loop contributions, which read
\bea
\Delta V^T = \frac{T^4}{2 \pi^2} \left[ \sum_b  g_b J_+ (m_b^2 / T^2) -
\sum_f  g_f J_- (m_f^2 / T^2) \right],
\eea
with the definitions
\bea
J_\pm(y^2) = \int_0^\infty dx \, x^2 \, \log( 1\mp \exp(- \sqrt{x^2 + y^2}) ).
\eea

Discussing a strong first-order phase transition in the MSSM, it is 
important to modify this expression by undergoing a two-step procedure, 
first deriving a 3D effective action and then treating this further
with two-loop perturbation theory~\cite{Bodeker:1996pc}  or, more 
safely,  with lattice numerical methods~\cite{Laine:2000rm}.  A 
simplified prescription, the above one-loop expression modified by "daisy" 
resummation, follows the same direction. It  allowed  the formulation of
the postulate of a "light" stop, in order to obtain a strong phase 
transition, in a transparent way~\cite{Moreno:1998bq}.
In the nMSSM model the strong first-order phase transition should be
triggered by the tree-level Lagrangian and we do not need this refined 
analysis. Just adding the "daisy" correction does not necessarily
improve the analysis. 

\section{Constraints at Zero Temperature}

Before we analyze the phase transition and compute the produced baryon asymmetry, we
confront the model with constraints coming from collider physics. 
The present limits are summarized in 
Tab.~\ref{tab_masses}.

In models with extended Higgs sectors, the Higgs couplings deviate
from the SM values. Of particular importance is the $ZZ\mathcal{H}_i$
vertex, where $\mathcal{H}_i$ denote the neutral Higgs mass
eigenstates, 
as computed from the one-loop potential. 
The size of this vertex is reduced by a factor 
\be
\xi_i =  \cos(\beta) \mathcal{O}_{i1}  + \sin(\beta) \mathcal{O}_{i2}, \label{xi_eq}
\ee
with respect to the SM value.
The matrix $\mathcal{O}_{ik}$ relates the mass eigenstates with
the two CP-even flavor eigenstates $S_1, S_2$.
In the CP-violating case, $\mathcal{O}$ is a $5 \times 5$ matrix without
block-diagonal structure, and the special form of Eq.~(\ref{xi_eq}) is due
to our convention for CP-even Higgs states (see Appendix~\ref{app_masses}).
If the neutral Higgs mass is below the value given in  
Tab.~\ref{tab_masses}, the LEP bound translates into an upper bound on $\xi$, as given
in Ref.~\cite{LEP}.

\begin{table}
\begin{tabular}{|c|c|}
\hline 
species & mass bound \\
\hline 
\hline 
charginos  $\tilde \chi^\pm $ & $\gtrsim 104$ GeV \\
\hline 
neutralinos $\tilde \chi^0$ & $\gtrsim 25$ GeV \\
\hline 
charged Higgses $H^\pm$ & $\gtrsim 90$ GeV \\
\hline 
neutral Higgses $H^0$  & $\gtrsim 114$ GeV \\
\hline 
\end{tabular}
\caption{Mass constraints on the spectrum\label{tab_masses}}
\end{table}

We do not implement any constraints on the squark spectrum, but choose the 
following stop mass parameters as used in Ref.~\cite{Menon:2004wv}:
\be
m^2_Q = m^2_U = 500 \textrm{ GeV}, \quad a_t = 100 \textrm{ GeV}.
\ee

The nMSSM suffers from a light singlino state. Because of the missing
singlet self coupling, this state acquires its mass only by mixing with the Higgsinos.
This is an important difference from more general singlet models, such as the
one discussed in Ref.~\cite{Huber:2000mg}. If this lightest neutralino has a mass
$m_{\tilde\chi^0} < m_Z/2$, it contributes to the invisible $Z$ width, leading to the
constraint \cite{LEP}
%
\be
\textrm{BR}(Z\to \tilde\chi^0 \tilde\chi^0 )
= \frac{g^2}{4\pi} \frac{(|U_{13}|^2 - |U_{14}|^2)^2}{24 \cos^2(\theta_W)}
\frac{m_Z}{\Gamma_Z} \left( 1- \frac{4m^2_{\tilde\chi^0} }{m^2_Z}\right)^\frac32
< 0.8 \times 10^{-3}. \label{Z-width}
\ee
Here, $U$ denotes the unitary matrix that diagonalizes the neutralinos as 
defined in Appendix~\ref{app_masses} and $\Gamma_Z = 2.5$ GeV denotes the $Z$ width.

A light neutralino can be avoided if the Higgs singlet coupling $\lambda$
is taken to be large. For large values of $\lambda$ a Landau pole is encountered 
below the GUT scale. Avoiding this Landau pole requires $\tan\beta>1.3$ and
$\lambda<0.8$ \cite{Menon:2004wv}, but we will also consider larger values of 
$\lambda$. This can be motivated by the so-called ``Fat Higgs'' models,
where the Higgs becomes composite at some intermediate scale \cite{HKLM03,DT05}.

In Ref.~\cite{Menon:2004wv} the model was further constrained by the
relic neutralino density. Here we will not impose this constraint directly.
However, from  Ref.~\cite{Menon:2004wv} we take the bound 
$m_{\tilde\chi^0}>25$~GeV,
which ensures that the dark matter density remains below the observed
value. For much larger masses the relic neutralino density is quite small,
so that neutralinos will only provide a fraction of the total dark matter in
the Universe.

\section{Electroweak Baryogenesis\label{ewbg}}

\subsection{Sources of CP violation in the nMSSM}

As discussed earlier, in the nMSSM there is the possibility of
additional CP violation in the singlet sector. As in the MSSM, the
relevant source in the transport equations usually comes from the
charginos, even though the neutralinos can in certain cases
contribute sizable effects as well~\cite{Cirigliano:2006dg}.  In the
interaction basis, where the Higgsinos and Winos are the
quasi-particles, their mass matrix takes the following form:
\bea
M_{\tilde\chi^\pm} = \begin{pmatrix}
0 & Y_{\tilde\chi^\pm}^T \\
Y_{\tilde\chi^\pm} & 0 \\
\end{pmatrix}, \quad
Y_{\tilde\chi^\pm} = \begin{pmatrix}
M_2 & g\phi_2(z) \rme^{-i\frac{q(z)}2} \\
g\phi_1(z) \rme^{-i\frac{q(z)}2} & \mu(z) \\
\end{pmatrix},
\eea
where $z$ denotes the direction along which a nearly planar bubble wall
is moving.
Unlike what happens in the MSSM, the effective $\mu$ term acquires a $z$-dependence
\be
\mu(z) = - \lambda \, \phi_s(z) \, e^{i \, q_s(z)}. 
\ee

The leading contribution (to the left-handed current) to second order
in the gradient expansion is proportional to~(see Eq. (92) in
Ref.~\cite{kps}, and Ref.~\cite{Kainulainen:2001cn})
\be S^{(2)} \sim  \left\{ m^{\dagger\prime\prime}m -m^\dagger m^{\prime\prime},
\partial_{k_z} \hat g^{\rm eq} \right\}^D, 
\label{sec_order_source} 
\ee
where $\hat g^{\rm eq}$ denotes the zero component of the vector part
of the chargino Green function in thermal equilibrium and in the
interaction basis.
The superscript $D$ indicates that the
diagonal entries in the mass eigenbasis are projected out following
the conventions of Ref.~\cite{kps}.  The first-order sources that are
used to calculate the BAU in the MSSM and nMSSM in Refs.~\cite{kps}
and~\cite{cmqsw,cmqsw2} become prominent when chargino mass
eigenstates are nearly degenerate.  Here these sources are generally
expected to be suppressed with respect to the second-order
source~(\ref{sec_order_source}), because we are mainly interested in the
generic non-degenerate case.  Note that gradient expansion applies
when the typical momentum of the particles is large with respect to the
inverse wall thickness, $k\gg \ell_w^{-1}$. Since $k\sim T$, this
condition is reasonably well satisfied even for rather thin walls
considered in this paper.  Therefore we expect that the sources that
are not captured by the gradient expansion 
Eq.~(\ref{sec_order_source}) --- an important example of which is
quantum mechanical CP-violating reflection~\cite{Cohen:1994ss} --- are
subdominant. This then suggests that the
source~(\ref{sec_order_source}) provides quite generically the main
contribution to the chargino-mediated BAU in the nMSSM.

In the MSSM, where $q(z)=0$ and $\partial_z \mu =0$,
 the evaluation of the second-order source~(\ref{sec_order_source}) leads 
to 
\be
S^{(2)}_{\rm MSSM} \sim  \left\{ 
g^2 \, \begin{pmatrix}
0 & M_2 \phi_2^{\prime\prime} -  \mu^* \phi_1^{\prime\prime}\\
 -M_2^* \phi_2^{\prime\prime} +  \mu \phi_1^{\prime\prime} & 0 \\
\end{pmatrix}
, \partial_{k_z} \hat g^{\rm eq} \right\}^D;
\ee
using the conventions of Ref.~\cite{kps}, this can be written as
\be
{\rm Tr}\, S^{(2)}_{\rm MSSM} \sim  g^2 \, \frac{\textrm{Im} (M_2 \mu)}{\Lambda}\, 
(\phi_1^{\prime\prime} \phi_2 + \phi_1\phi_2^{\prime\prime})\,
 \partial_{k_z} \hat g^{{\rm eq} , L}, 
\label{MSSMsource}
\ee
where $\Lambda$ denotes the difference of the eigenvalues of the matrix $m^\dagger m$.

In the nMSSM, there are various additional contributions from the
derivatives acting on $\mu$ in the source
(\ref{sec_order_source}), especially a novel diagonal term of
the following form:
\be
S^{(2)}_{\rm nMSSM} \sim  \left\{ 
 \, \begin{pmatrix}
0 & 0 \\
0 & \mu^{\prime\prime *} \mu -  \mu^* \mu^{\prime\prime} \\
\end{pmatrix}
, \partial_{k_z} \hat g^{\rm eq} \right\}^D.
\label{newsource}
\ee
This contribution dominates if $\mu \ll M_2$, which  is
usually the case in the nMSSM since $\mu$ is related to the singlet vev, 
while the Wino mass parameter $M_2$ is not related to the
parameters of the singlet sector and expected to be of the SUSY scale.

Hence, we will consider two scenarios. First, we consider the case of
large $M_2 \approx 1$~TeV. In this region the
contribution~(\ref{newsource}) will almost coincide with the full
expression (\ref{sec_order_source}). Second, we will choose a rather
small Wino-mass parameter $M_2
\approx 200$ GeV.  In this case the additional contributions in
Eq.~(\ref{sec_order_source}) can lead to an enhancement or a cancellation
in the BAU, and one should keep in mind that the neglected
mixing effects could contribute as well. In both cases, using the full
second-order source~(\ref{sec_order_source}), the baryon to entropy
ratio $\eta$ is determined as was done in Ref.~\cite{kpss} for the MSSM.

We use a system of diffusion equations that was first derived in
Ref.~\cite{Huet:1995sh} and later adapted in Refs.~\cite{JCK,
Carena:1997gx,cmqsw, cmqsw2,kpss,FH06}. This system describes how the
CP violation is communicated from the chargino sector to the
left-handed quarks and finally biases the sphaleron processes. These
diffusion equations rely on certain assumptions, e.g.~that the
supergauge interactions are in equilibrium, and hence lead to sizable
uncertainties. Furthermore, we do not take into account recent
developments in the determination of interaction rates presented in
Refs.~\cite{Lee:2004we, Cirigliano:2006wh}, but employ the 
parameters of Ref.~\cite{kpss}.  Nevertheless, the accuracy
of the determined BAU should be sufficient for the analysis in this work.

In the following we will briefly focus on the term~(\ref{newsource}),
which is prominent in the limit of large $M_2$.  Beside the critical
temperature $T_c$, the generated BAU only depends on
the profile of the Higgsino mass parameter $\mu(z)$ during the phase
transition, namely the change of the phase $\Delta q_s$, the wall
thickness $l_w$ and the profile of $|\mu (z)|$.

\begin{figure}[htbp]
\begin{center}
\epsfig{file=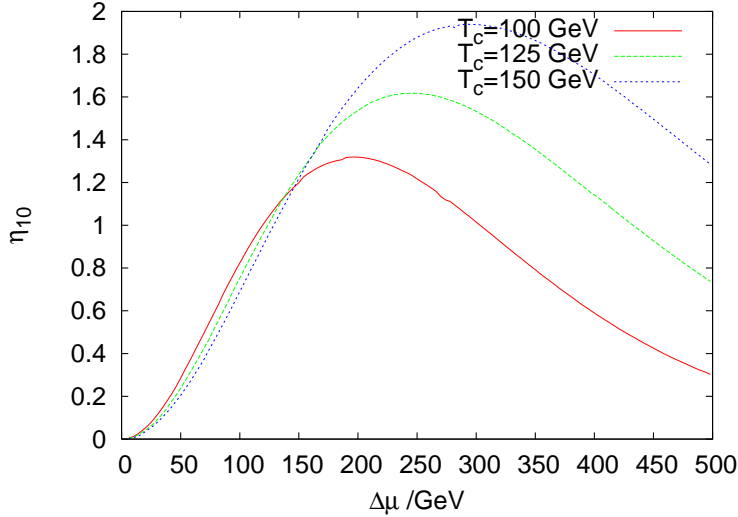, width=4.0 in}
\end{center}
\vskip -0.2in
\caption[Fig:fig_eta]{%
\label{fig_eta}
\small
The dependence of $\eta_{10} \equiv 10^{10}\eta$ 
on the chargino mass parameter $\Delta\mu=\mu_0$.
The parameters used are $l_w = 10/T_c$ and $\Delta q_s=\pi/10$.
}
\end{figure}

With good accuracy, the baryon to entropy ratio,
$\eta  \equiv {n_B}/{s}$, scales as 
\be
\eta \propto \frac{\Delta q_s}{l_w \, T_c}. \label{eta_scale}
\ee
The dependence on the profile of $|\mu(z)|$
is shown in Fig.~\ref{fig_eta}. In this example, the profile 
is parametrized by
\bea
|\mu(z)| &=& \mu_0 - \Delta \mu \left( \frac12 + \frac12 \tanh{(z / l_w)}  \right) \\
q_s(z) &=& \Delta q_s \left( \frac12 + \frac12 \tanh{(z / l_w)}  \right),
\eea
and the values $\mu_0=\Delta\mu$, $l_w = 10/T_c$,
$\Delta q_s=\pi/10$ and several values of $T_c$ have been chosen. 

To give some feeling about the produced BAU, we note that a good
estimate of the predicted $\eta_{10}$ is in this case given by the
formula
\be
\eta_{10} \approx \, c(T_c) \,\,\frac{\Delta q_s}{\pi} \frac{1}{ \, l_w \,T_c } \,  
\left( \frac{\mu_0}{\tau T_c} \right)^{\frac32} \, \frac{\Delta\mu}{\tau T_c} \,
\exp(-\mu_0 / \tau T_c ),
\ee
with $c(T_c)\approx 1.6 \, T_c/$GeV and $\tau\approx 0.78$.
This formula characterizes the BAU in the case of large $M_2 \approx 1$~TeV
and $\mu_0 \approx \Delta \mu$.

\subsection{EDM constraints in the nMSSM}

The most severe experimental constraints on CP violation come from
measurements of the EDM of the electron,
$d_e < 1.6 \times 10^{-27} \,e \, {\rm cm}$ \cite{Regan:2002ta},
and neutron, $d_n < 3.0 \times 10^{-26} \,e \, {\rm cm}$ \cite{Baker:2006ts}.
Already at the one-loop level, contributions of the
superpartners give sizable effects in the case of CP-violating phases of
 $O(1)$.  These one-loop diagrams are the same in the nMSSM and
MSSM, setting $\mu=-\lambda \langle S \rangle$. Minimizing the Higgs
potential, we can compute the phase of the effective $\mu$
parameter. Of course, this phase could be neutralized by introducing
a compensating phase in the parameter $M_2$.  Such a tuning would
allow us to eliminate the one-loop EDMs completely, without much affecting
the generated baryon asymmetry, since the dominating
source is proportional to the change in the phase $\Delta q_s$ and not
sensitive to the value of $q_s = \arg(\mu)$ in the broken phase.  This
is not possible in the MSSM, because there the produced BAU is, like the
electron EDM contribution from the charginos, proportional to the
combination $ {\rm Im} (\mu M_2) $.

However, here we take a different approach, using only the phase $q_t$
in the Higgs potential as sole source of CP violation. The one-loop
EDMs then induce mass bounds for the first and second generation
squarks and sleptons; depending on the model parameters these are in the
range from a few TeV up to $50$ TeV~\cite{Pospelov:2005pr}. Since the
constraints from the neutron EDM are usually less stringent than the
ones coming from the electron EDMs, we will focus in the following on
the latter.  In single cases we also calculated the Barr--Zee type
contributions to the neutron EDM~\cite{Barr:1990vd}, but they barely
reach the most recent experimental bounds~\cite{Baker:2006ts}.

Besides heavy sfermion masses, there is another possibility to
suppress the one-loop electron EDM in our model. Notice that the
absolute value of the CP-violating phase $(q_s+q)$ can be smaller in
the broken phase than in the symmetric phase. This phase $(q_s+q)$ is
the only CP-odd combination that enters the electron EDMs on the
one-loop level in our simplified nMSSM model, with CP violation only in
the $t_s$ parameter.  Hence, there is the possibility that today's
observed $\sin(q_s+q) \ll 1$, even though the phase $q_s$ greatly
changed during the phase transition thus producing the BAU. We will
analyze this possibility in detail in Sec.~\ref{numana}. The explicit
form of the one-loop electron EDM contributions is given in
Appendix~\ref{app_edm}. This possibility entails a certain amount of
tuning.

Additional EDMs can be generated from two-loop chargino or Higgs
graphs (see, for instance, \cite{Pospelov:2005pr} and references therein). 
Notice that the MSSM two-loop chargino contribution to the electron
EDM~\cite{Chang:2002ex,Pilaftsis:2002fe} is proportional to $\tan
\beta$ and hence subleading in our model that usually predicts $\tan
\beta \sim O(1)$.  Potentially
harmful diagrams, including the additional CP-odd scalar in the nMSSM, are
small as well because of the modest $\tan \beta$ and because
only the Higgs component of the CP-odd scalars couples to the
charginos, while the singlet component delivers the additional
CP violation~(for a calculation of these contributions see
\cite{Chang:1998uc}).

\section{Electroweak Phase Transition\label{ewpt}}

  One of the parameters entering our baryogenesis analysis is the thickness 
of the Higgs wall profile during the electroweak phase transition $l_w$. 
Since our CP-violating source is a second-order effect in gradients,
the integrated BAU scales as $\eta \sim 1/l_w$, as already mentioned in
 Eq. (\ref{eta_scale}).

To determine the wall thickness $l_w$ one has to examine the dynamics
of the phase transition~\cite{MooreProkopec:1995}. This has been done
for the MSSM in Ref.~\cite{Moreno:1998bq} and for the NMSSM in
Ref.~\cite{Huber:2000mg}. Typical values for the MSSM seem to be close
to $l_w=10/T_c$.  In the nMSSM, we expect rather thin wall profiles,
since the linear singlet term and the trilinear singlet Higgs term in
the effective potential will make the phase transition much stronger
than the loop-suppressed stop corrections that are
responsible for the first-order phase transition in the MSSM. In the
light of Eq. (\ref{eta_scale}) this will further enhance the produced
asymmetry with respect to the MSSM case.

To determine the dynamical parameters of the wall, we solved for the classical
bounce
solution of the Higgs and singlet fields $(\phi_1, \phi_2, \phi_s, q, q_s)$ at the critical
temperature (where the two minima of the potential are degenerate).
Our numerical approach is based on the variation of the classical action 
and is discussed in detail in Ref.~\cite{Konstandin:2006nd}.

\begin{figure}[htbp]
\begin{center}
\epsfig{file=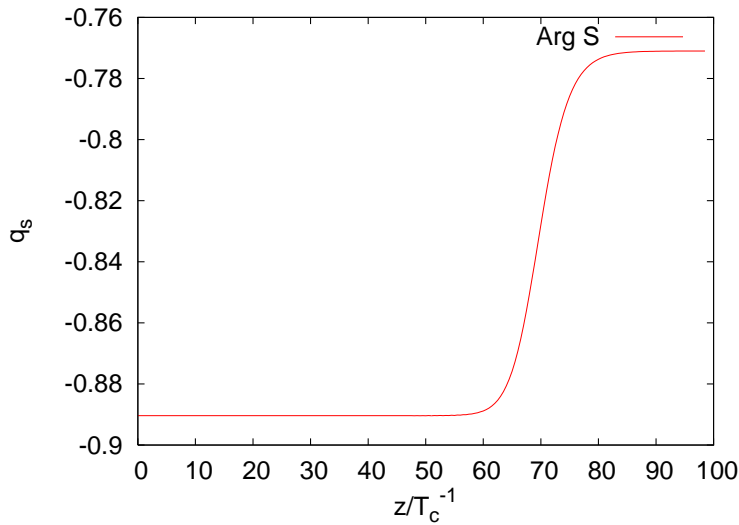, width=4.0 in}
\end{center}
\vskip -0.2in
\caption[Fig:fig_qs]{%
\label{fig_qs}
\small
A typical wall profile for the parameter $q_s$,
corresponding to the parameters $\Delta q_s=0.119$ and $l_w = 4.81\, T_c^{-1}$.
}
\end{figure}

Another parameter relevant to the dominating source in
Eq.~(\ref{newsource}) is the profile of the CP-violating phase of the
singlet field. A typical solution is displayed in Fig.~\ref{fig_qs}
corresponding to the parameters $\Delta q_s=0.119$ and $l_w = 4.81 \,
T_c^{-1}$.

To illuminate a little bit the nature of the phase transition, we will
recall some approximate analytical criterion for the occurrence of a first-order
phase transition, first given in a slightly
different way in Ref.~\cite{Menon:2004wv}. Consider the tree-level
potential in Eq. (\ref{tree-pot}) without CP violation, $q_t
=0$. Additionally assume that the temperature effects give rise to
the effective potential

\be
\Delta V^T = \alpha \, \phi^2 T^2,
\ee
where $\alpha$ is some unspecified positive constant. 
In Ref.~\cite{Menon:2004wv} it was shown that a necessary condition
for a first-order phase transition is approximately given by

\be
m_s^2 < \frac{1}{\tilde \lambda} \left| \frac{\lambda^2 t_s}{m_s} - m_s \tilde a \right|.
\label{pt_con}
\ee%

This can be seen in the following way.
Given a certain value for $\phi$, $\phi_s$ can easily be evaluated to
be 

\be
\phi_s = -\left( \frac{t_s + \tilde a \phi^2}{m_s^2+ \lambda^2 \phi^2} \right).
\ee

Using this in our potential, and expanding around $\phi=0$, we obtain
\bea
V + \Delta V^T &=& 
-\frac{t_s^2}{m_s^2} +  \left( M^2 + \alpha \, T^2 \right) \phi^2
+ \tilde \lambda^2 \, \phi^4 - \frac{(t_s + \tilde a \phi^2)^2}{m_s^2+ \lambda^2 \phi^2}
\nn \\
&=& c_0 + c_1 \phi^2 + c_2 \phi^4 + c_3 \phi^6 \cdots,
\eea
with the coefficients%
\bea
c_0 &=& -\frac{t_s^2}{m_s^2}, \nn \\
c_1 &=& M^2 + \alpha \, T^2 - \frac{2 \tilde a t_s}{m_s^2} 
+ \frac{\lambda^2 t_s^2}{m_s^4}, \nn \\
c_2 &=& \tilde \lambda^2 - \frac{1}{m_s^2} 
\left( \tilde a- \frac{\lambda^2 t_s}{m_s^2}\right)^2, \nn \\
c_3 &=& \frac{\lambda^2}{m_s^4} \left( \tilde a- \frac{\lambda^2 t_s}{m_s^2}\right)^2.
\eea
If the symmetric minimum is absent at zero temperature, $c_1(T=0)<0$,
a temperature $T_2$ can be found, such that $c_1(T_2)=0$.  For this
temperature there exists a lower-lying potential minimum in the case
$c_2<0$, which is equivalent to the condition in Eq.~(\ref{pt_con}).
Since for temperatures $T\gtrsim T_2$ a potential well develops
between the symmetric and the lower broken vacuum, a first-order phase
transition is possible, given the vacuum decay rate is large enough
such that the transition occurs before the temperature $T_2$ is
reached. Hence, it is possible in the nMSSM to obtain a first-order
phase transition due to tree-level dynamics, in contrast to the MSSM.
Analyzing the numerical results, we will see that the constraint in
Eq.~(\ref{pt_con}) is usually fulfilled in viable models, even if the
one-loop contributions to the potential and the CP phase are included
and hence that the phase transition is dominated by the tree-level
dynamics.

The argument just presented is a concrete realization of the general effective
field theory approach recently discussed in Refs.~\cite{GSW04,BFHS04}. 
There it was shown that
a strong first-order phase transition can be induced at tree-level by the 
interplay of a negative $\phi^4$ term
and a positive $\phi^6$ term which stabilizes the potential.
The suppression scale of the $\phi^6$ term should be somewhat below a TeV
for the mechanism to work. Here the relevant $\phi^4$ and $\phi^6$ operators
are generated by integrating out the singlet field.
This generalizes the usual situation, where
the phase transition is induced by a negative  $\phi^3$ term and a positive
$\phi^4$ term.

\section{Numerical Analysis\label{numana}}

To inspect the parameter space of the nMSSM we proceed as
follows. First, we choose random parameters for the Higgs potential in
the ranges displayed in Tab.~\ref{tab_param}. To ensure maximal
numerical stability, all chosen parameters are of  $O$(1), and can
be thought of as dimensionless parameters.  Those parameters then lead
not to the physical Higgs vev, but to some dimensionless Higgs
vev $\phi_0$. Finally, all dimensionful quantities, such as the
critical temperature or the mass spectrum have to be scaled with
$(173.458$ GeV)/$\phi_0$ to yield the physical values. During
the minimization of the potential, stable and metastable broken minima
were analyzed. Depending on the parameters, metastable minima
occur but, in the CP-conserving case $q_t=0$, no transitional
(spontaneous) CP violation was observed in contrast to the
NMSSM~\cite{Huber:2000mg} that contains an additional cubic singlet term.
 
Next, we correct the bare parameters with
the one-loop contributions of Eq.~(\ref{C-W-log}) and confront them
with the constraints on the mass spectrum from Tab.~\ref{tab_masses}
and on the Z-width from Eq. (\ref{Z-width}).

\begin{table}
\begin{tabular}{|ccccc|}
\hline 
\,\,\, lower bound  && parameter && upper bound \,\,\, \\
\hline 
\hline 
0.1 &$<$& $m_1$ &$<$& 1 \\ 
\hline 
0.1 &$<$& $m_2$ &$<$& 1 \\ 
\hline 
0.1 &$<$& $m_{12}$ &$<$& 2 \\ 
\hline 
0 &$<$& $m_s$ &$<$& 2 \\ 
\hline 
$-2$ &$<$& $\lambda$ &$<$& 2 \\ 
\hline 
0 &$<$& $a_\lambda$ &$<$& 2 \\ 
\hline
0 &$<$& $t_s^{1/3}$ &$<$& 2 \\ 
\hline 
0   &$<$& $q_t$ &$<$& $2\,\pi$ \\ 
\hline 
\end{tabular}
\caption{Dimensionless parameter ranges used for the numerical analysis.
\label{tab_param}}
\end{table}

If the parameter set passes these constraints, we add the temperature
dependent contributions to the effective potential as explained in
Sec.~\ref{temp} and examine the phase transition. We require that the
models have a first-order phase transition of sufficient
strength~\cite{Menon:2004wv,Shaposhnikov:1987pf}, $\phi_0 / T_c >0.9$.

Before we discuss baryogenesis in our model, we would like to examine
restrictions on the parameters imposed by the constraints on the mass
spectrum and comment on the criterion for a first-order phase
transition given in the last section in Eq.~(\ref{pt_con}).

First, the eight parameters given in Tab.~\ref{tab_param} have to lead
to the correct Higgs vev, which is achieved by a rescaling of the
dimensionful parameters. Hence our parameter space is effectively only
seven dimensional. One restriction on the parameters is that 
$a_\lambda$ cannot be chosen arbitrarily large, since this
destabilizes the potential in the negative $\phi_s$ direction.
Analyzing the parameter sets that fulfill the mass constraints,
one observes that the parameters  $\lambda$ and $q_t$  are not distributed
homogeneously. Small values of
$\lambda$ make it seemingly difficult to fulfill the mass bound of
the chargino, since one of the diagonal entries of the chargino mass
matrix is $-\lambda \phi_s$. To have a potential with extremely large
vev $\phi_s$ requires at least some fine-tuning, since the
one-loop contribution tends to yield an effective potential that is
unbounded from below, if the tree-level parameters are chosen to provide
a large vev $\phi_s$. Large values of $\lambda$ hence seem to be
the more natural choice, even though they can lead to a Landau
pole~\cite{Menon:2004wv}. Usually, the mass constraints on the
neutralinos are automatically fulfilled, if the charginos
surpass their more restrictive bounds, but additional constraints on
the parameters enter through the spectrum of the Higgs
particles. In many cases, a range of values for the parameter $a_\lambda$ can be found,
where off-diagonal elements in the Higgs mass matrix cancel, which tends to
enlarge the lightest Higgs mass. In addition the parameter $a_\lambda$ has
a strong influence on the phase transition according to Eq.~(\ref{pt_con}).

\begin{figure}[t]
\begin{center}
\epsfig{file=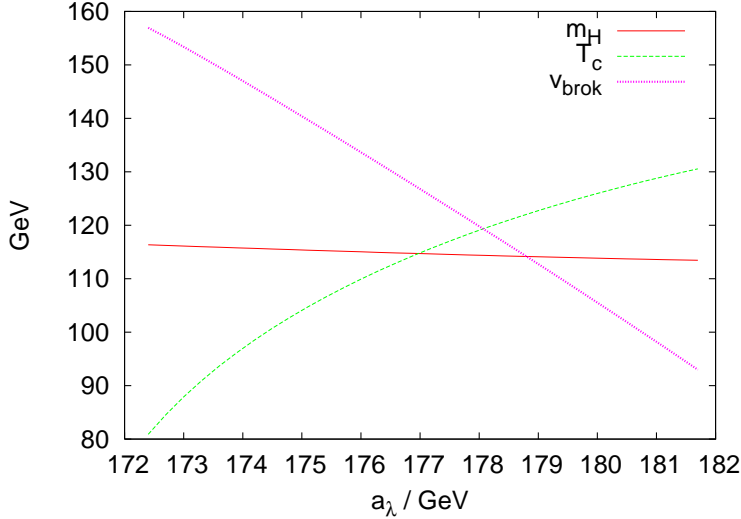, width=4.0 in}
\end{center}
\vskip -0.2in
\caption[Fig:al]{%
\label{fig_al}
\small
The critical temperature $T_c$, the Higgs vev $\phi$ in the broken phase
at $T_c$ and one Higgs mass as functions of $a_\lambda$.}
\end{figure}

This situation is demonstrated in Fig.~\ref{fig_al} for a parameter set with a
rather small parameter $\lambda$. The parameters
$m_1,m_2,m_{12},m_s$ are chosen such that $\tan(\beta)=2.0$,
$\phi_s=-250$ GeV, $\phi=173$ GeV and $M_a=500$ GeV at the tree-level,
where the CP-odd Higgs mass parameter is defined by
\be
M_a^2 = \frac{1}{\sin{\beta}\cos{\beta}} \left( m^2_{12} - a_\lambda \phi_s\right).
\ee
The remaining parameters are $\lambda=0.55$, $t_s^{1/3} = 70$
GeV, $q_s=0.3$, while $a_\lambda$ is varied. For $172$ GeV $<
a_\lambda < 178$ GeV, this model develops a strong first-order phase
transition and generates more than the observed BAU. 
For lower values of $a_\lambda$, the model has no stable broken phase, while 
for larger values of $a_\lambda$, the phase transition is too weak. The plotted
Higgs mass is that of the third lightest Higgs, but the two lighter states
would have escaped detection at LEP because of  the suppressed coupling to the
$Z$-boson. This example demonstrates that even though there exist viable models
without Landau pole in $\lambda$, this possibility entails a
certain amount of tuning.

\begin{figure}[t]
\begin{center}
\epsfig{file=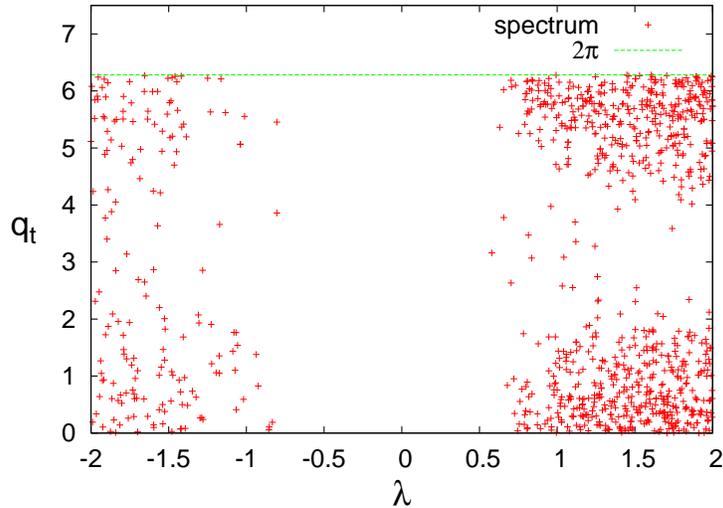, width=4.0 in}
\end{center}
\vskip -0.2in
\caption[Fig:lq_check]{%
\label{fig_lq_check}
\small
The parameters $\lambda$ and $q_t$ for a set of random models
that fulfill the mass constraints. }
\end{figure}

The second parameter that is restricted by
the mass constraints is the CP-violating phase $q_t$. The reason for
this effect is that values with $\cos(q_t) \approx -1$ lead to
smaller Higgs masses. Fig.~\ref{fig_lq_check} displays the parameters
$\lambda$ and $q_t$ for a set of random models that fulfill the mass
constraints.

\begin{figure}[b]
\begin{center}
\epsfig{file=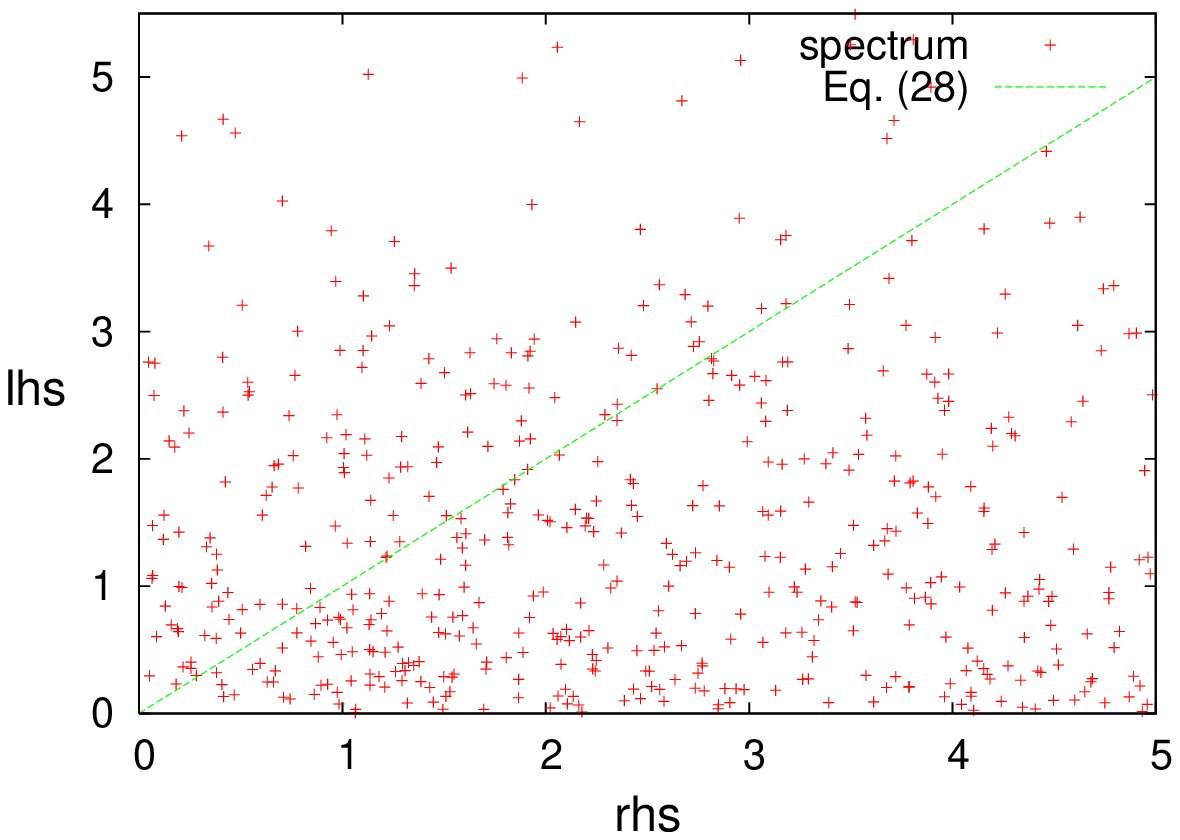, width=3.2 in}
\epsfig{file=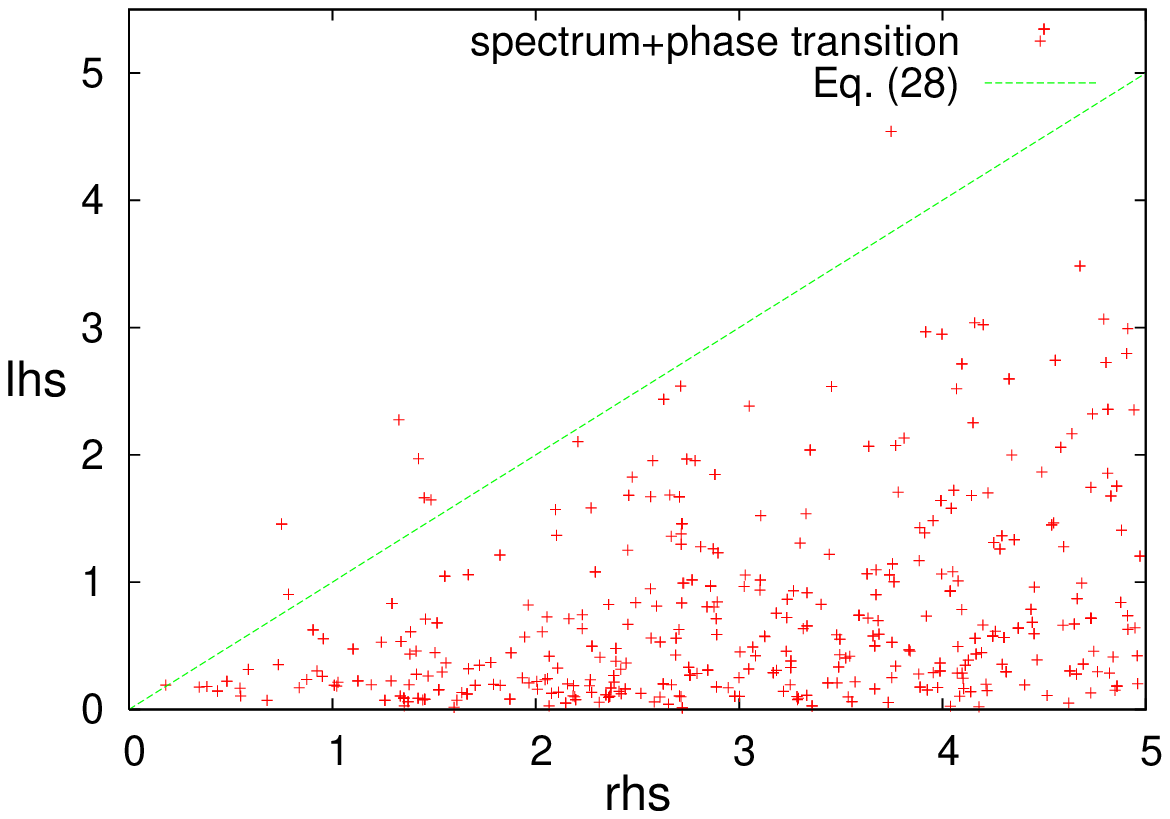, width=3.2 in}
\end{center}
\vskip -0.2in
\caption[Fig:pt_check]{%
\label{fig_pt_check}
\small
The plots show the combinations of the parameters that enter the
tree-level condition for a first-order phase transition,
Eq.~(\ref{pt_con}).  The left plot contains parameter sets that 
fulfill only the mass constraints, while the right plot contains
parameter sets that have in addition a strong first-order phase transition.}
\end{figure}

Demanding a strong first-order phase transition further restricts the
parameter space.  In Fig.~\ref{fig_pt_check} we plot the left-hand side
versus the right-hand side of the criterion in Eq.~(\ref{pt_con}) (both
sides are scaled by $1/(m_1 m_2)$ to make them dimensionless).  In the
left plot we use random models that fulfill the mass constraints on
the spectrum, but are unconstrained otherwise; in the right plot we
impose the mass constraints on the model and require a strong first-order phase transition.
In the latter case, most of the parameter sets are in
accordance with Eq.~(\ref{pt_con}), while the parameter sets in the
former case are evenly distributed. Hence, the tree-level criterion for the phase
transition seems to be applicable even if the one-loop contributions to
the effective potential and the CP phase are taken into account.

\begin{figure}[t]
\begin{center}
\epsfig{file=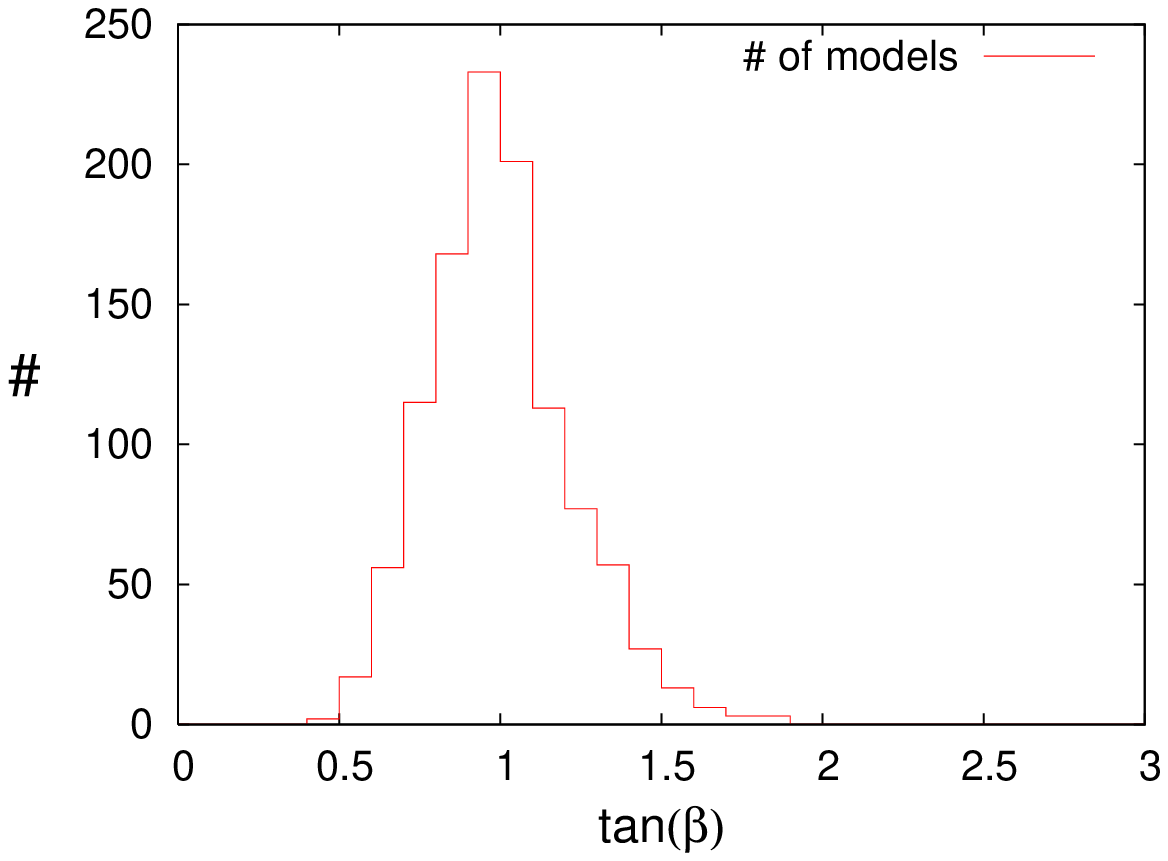, width=3.5 in}
\end{center}
\vskip -0.2in
\caption[Fig:fig_beta]{%
\label{fig_beta}
\small
The binned result of the parameter $\tan(\beta)$ for the parameter sets 
with a strong first-order phase transition.  }
\end{figure}

\begin{figure}[b]
\begin{center}
\epsfig{file=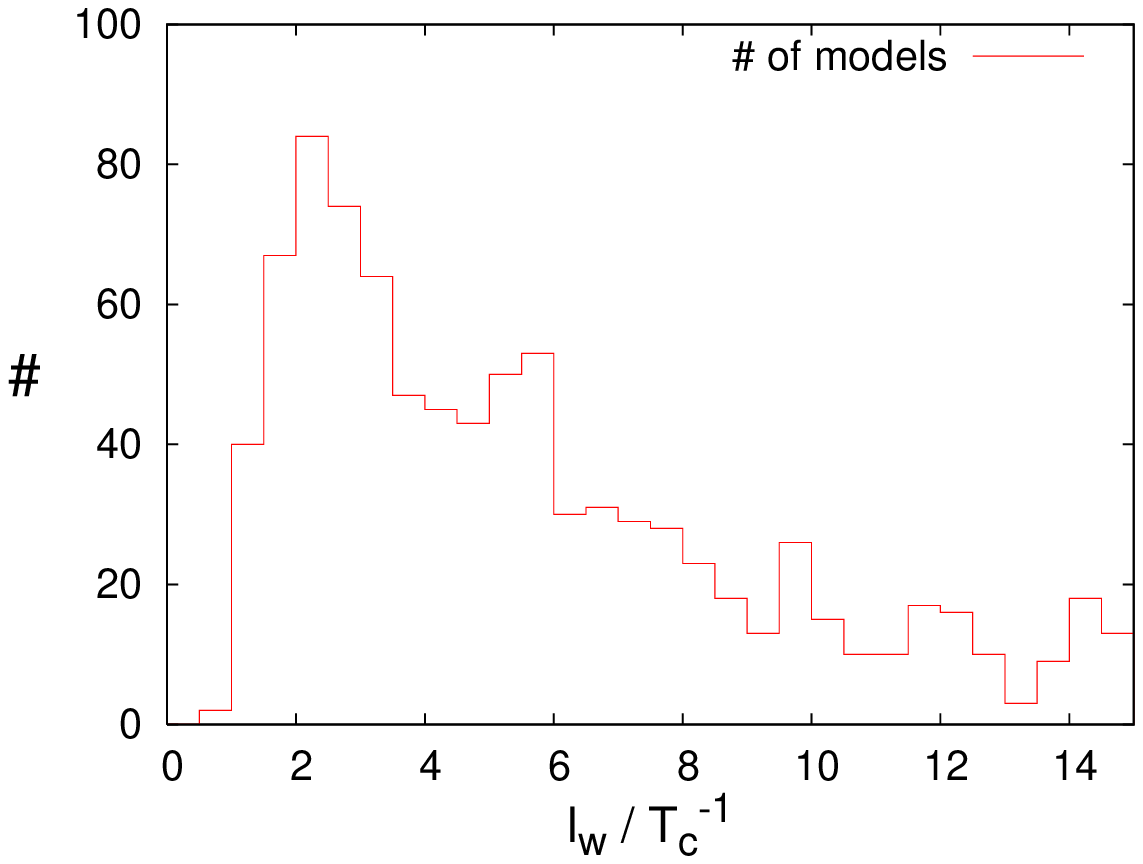, width=3.5 in}
\end{center}
\vskip -0.2in
\caption[Fig:fig_lw]{%
\label{fig_lw}
\small
The binned result of the parameter $l_w$ for the parameter sets 
with a strong first-order phase transition.
}
\end{figure}

In the nMSSM, for several reasons, we expect a much larger BAU than in the MSSM.
First, the parameter $\tan(\beta)$, which needs to be
large in the MSSM, is naturally $O(1)$ in the nMSSM as
depicted in Fig.~\ref{fig_beta}.  This does not only help to suppress the
two-loop contributions to the EDMs, as discussed in the previous section,
but also enhances the contributions from the
source in Eq.~(\ref{MSSMsource}).
Secondly, the wall thickness, which in the MSSM usually is of order
$20/T_c$--$30/T_c$~\cite{Moreno:1998bq}, can be much smaller, since the
phase transition is strengthened by the  linear and trilinear terms
 in the effective potential. This claim is supported by
Fig.~\ref{fig_lw}.
The third reason for the enhancement of the BAU in the nMSSM with respect
to the MSSM is the additional source in Eq.~(\ref{newsource}), which is in many
cases dominating the generation of the BAU.

\begin{figure}[t]
\begin{center}
\epsfig{file=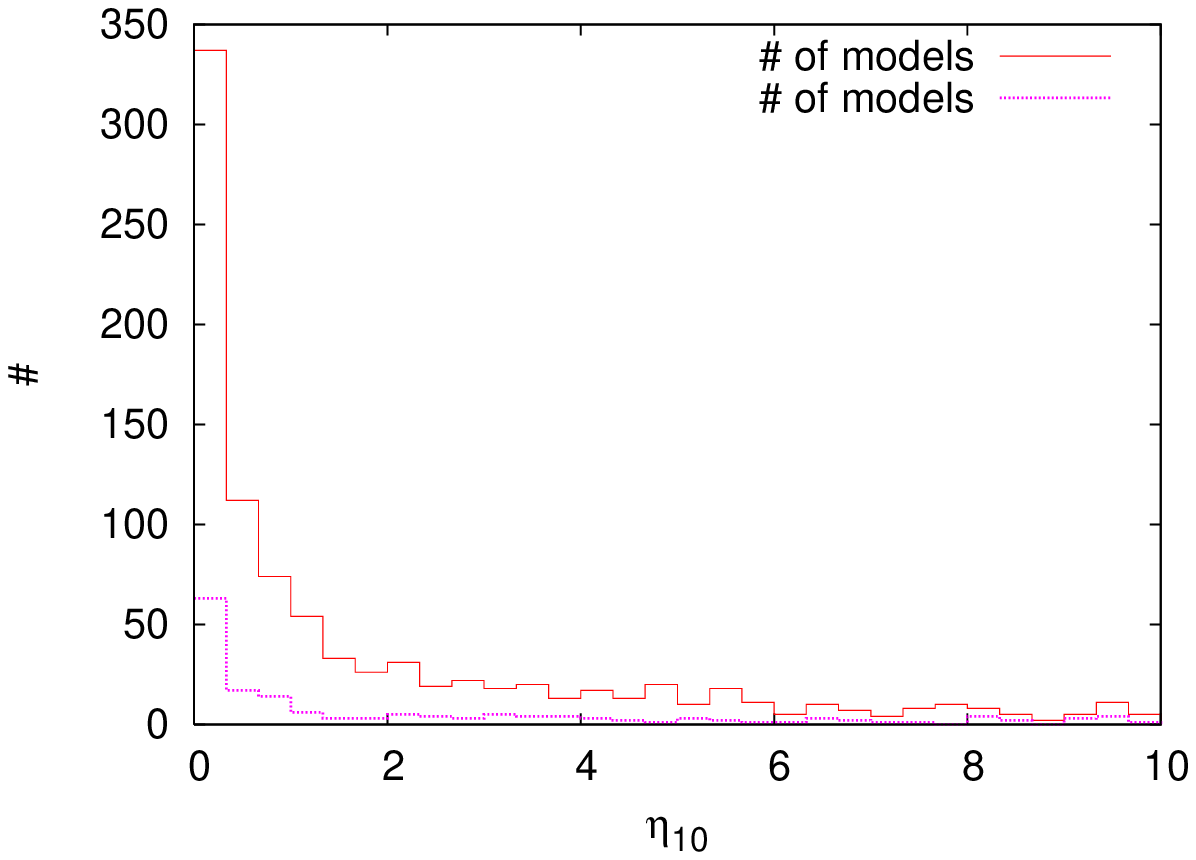, width=3.5 in}
\end{center}
\vskip -0.2in
\caption[Fig:fig_ewbg1]{%
\label{fig_ewbg1}
\small
The binned result for the BAU analysis for large $M_2 = 1$ TeV. 
Approximately 50\% of the parameter sets predict a value of the
baryon asymmetry higher than the observed one. The bottom line corresponds to
parameter sets that fulfill current bounds at the electron EDM with 1 TeV sfermions (4.8\%).
}
\end{figure}

\begin{figure}[b]
\begin{center}
\epsfig{file=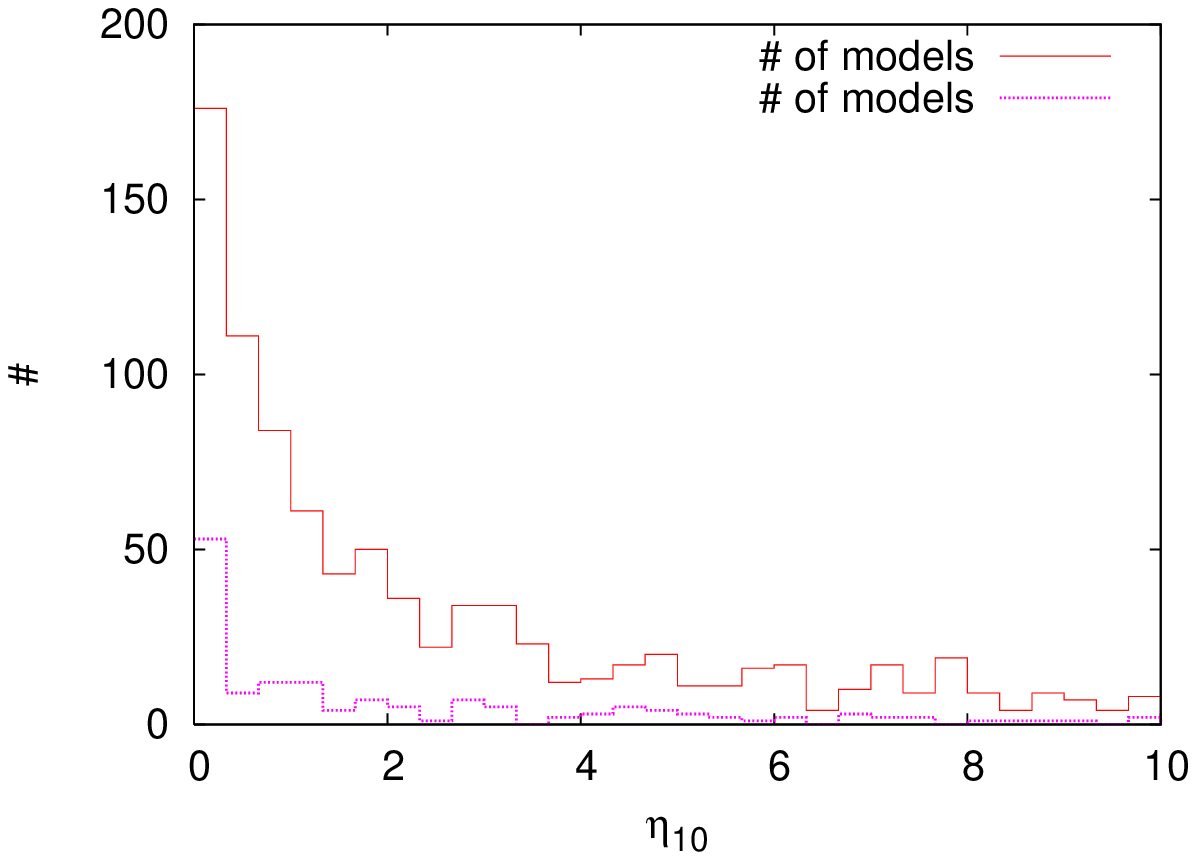, width=3.5 in}
\end{center}
\vskip -0.2in
\caption[Fig:fig_ewbg2]{%
\label{fig_ewbg2}
\small
The binned result for the BAU analysis for small $M_2 = 200$ GeV. 
Approximately 63\%  of the parameter sets predict a value of the
baryon asymmetry higher than the observed one. The bottom line corresponds to
models that fulfill current bounds at the electron EDM with 1 TeV sfermions (6.2\%).
}
\end{figure}

Finally, we calculate the generated baryon asymmetry and compare the result with the
experimental observation, $\eta=(0.87\pm0.03)\times10^{-10}$ \cite{eta}.
The result is shown 
in Figs.~\ref{fig_ewbg1} and \ref{fig_ewbg2}.
Approximately 50\% of the parameter sets predict a higher value than the observed baryon 
asymmetry
in the model with large $M_2=1$ TeV, while this number increases to 63\% for small
$M_2=200$ GeV.

In addition, we plotted the BAU generated by parameter sets that fulfill the
experimental bounds on the electron EDM with sfermion masses of
1 TeV in the first and second generation.  Some of them predict a BAU
in accordance with observation, and hence give the possibility to construct
nMSSM models that contain less constrained  sfermions (lighter than 1
TeV) being at the same time consistent with EDM constraints and
baryogenesis. In some cases the electron EDM is small because of a random
cancellation between the neutralino and chargino contributions, but
occasionally the suppression of the electron EDM is due to the fact
that the combination $\sin(q_s + q)$ is relatively small in the broken
phase.

\section{Conclusion\label{concl}} 

 We have analyzed the phase transition and baryogenesis in the 
nMSSM~(\ref{W_nMSSM}) -- (\ref{V_soft}) with CP violation in the singlet sector. 
We have shown that the singlet field enhances the strength of the phase transition
in such a way that one typically obtains a strong phase transition,
as required for successful baryogenesis.
This is to be contrasted with the MSSM, in which the mass of the 
lightest Higgs field must not be greater than about
$120~{\rm GeV}$, and the right-handed stop must be light,
$m_{\tilde t_R} \in\left(120,160\right)~{\rm GeV}$.

Next we performed the calculation of baryogenesis mediated by
charginos in the nMSSM. After calculating the CP-violating sources in
the gradient expansion, we argued that in most of the parameter space
the dominant source comes from the second-order semiclassical force in
the Boltzmann transport equation for charginos. The source related to flavor mixing,
of first order  in the gradient expansion, tends to be smaller, because the bubble 
wall is rather thin. For a generic choice of parameters, one is far from the
chargino mass degeneracy, where the first-order source may be
important.
To come to this conclusion, we used an approach to the calculation of
the first-order sources~\cite{kps,kpss} that differs from earlier
work~\cite{Menon:2004wv,cmqsw,cmqsw2} in the sense that our treatment
of sources is basis independent, and the magnitude of the source in
the transport equation is unambiguous. Using this advanced transport
theory, the first-order sources are of a somewhat lower amplitude and
exhibit a much narrower resonance near the chargino mass degeneracy,
with the effect that in most of the parameter space the second-order
source dominates. In the MSSM, this is not the case for the chargino-mediated 
baryogenesis because the wall tends to be thicker, thus weakening the
second-order (semiclassical force) source, while leaving the
first-order source more or less unchanged. Furthermore, the dominant
second-order source of baryogenesis in the nMSSM,
Eq. (\ref{newsource}), is not present in the MSSM. Owing to these
differences, successful baryogenesis in the MSSM is only possible near
the resonance (chargino mass degeneracy), and with nearly maximum CP
violation, which is in conflict with the current EDM bounds, unless a
tuning of parameters is invoked~\cite{kpss}.

On the other hand our analysis of the baryon production in the nMSSM
looks promising. When we restrict the CP-violating phase in the
singlet sector to be about $q_t \sim 0.3$, and take $\tan(\beta)\sim
1$, we still get baryon production consistent with the observed
value. This choice certainly does not violate any of the current EDM
bounds. When $q_t$ is chosen randomly, approximately $50\%$ ($63\%$)
models predict more than the observed BAU when $M_2=1~{\rm TeV}$
($M_2=200~{\rm GeV}$), which indicates that baryogenesis in the nMSSM
is generic.

It is finally interesting to compare baryogenesis in the 
nMSSM and in the general NMSSM formerly analyzed in Ref.~\cite{Huber:2000mg}.
Because of the presence of a singlet self coupling and an explicit $\mu$ term,
the NMSSM allows for a much richer Higgs phenomenology. (However,
some additional assumptions about the structure of the higher dimensional
operators have to be made, in order to prevent the destabilization of the
electroweak scale by corrections to the singlet tadpole.)
There is no danger
of a light singlino state, so that the model can account for the observed
baryon asymmetry also for values of $\lambda \lesssim 0.1$ and large $\tan\beta$. 
It also allows for transitional CP violation, which means an electroweak
phase transition that connects a high-temperature CP-broken phase with
a low-temperature CP-symmetric phase. This way there are varying complex
phases in the bubble wall, without leaving a trace at low temperatures.
In the present model this possibility is prevented by the strongly constrained
Higgs potential. 

In Ref.~\cite{Huber:2000mg} the source terms were computed by solving
the Dirac equation for the charginos in the WKB approximation. As has
been recently shown in Ref.~\cite{FH06}, this formalism can reproduce
the second-order source of Ref.~\cite{Kainulainen:2001cn}, used in the
present work, if the Lorentz transformation to a general Lorentz frame
is done carefully (see also Ref.~\cite{Kainulainen:2002th}).  These
results suggest that in Ref.~\cite{Huber:2000mg} the baryon asymmetry
was underestimated by a factor of 2 to 5. It seems to be interesting to
update this analysis to meet the present experimental constraints.  It
also seems promising to apply the presented techniques to more general
supersymmetric models, such as models with extra
U(1) symmetries \cite{Kang:2004pp}.

\section*{Acknowledgements}
We would like to thank M. Laine for discussing some subtleties of the
electroweak phase transition in the MSSM with us.  T.~K. is supported by the
Swedish Research Council (Vetenskapsr{\aa}det), Contract
No.~621-2001-1611.

\appendix

\section{Mass Spectrum of the nMSSM \label{app_masses}}

In the following we collect the mass matrices that have been used in the 
one-loop potentials in Sec.~\ref{model}. 

We used the physical constants
\bea
\sin(\theta_W) = 0.2312, &&\quad 
\alpha_{EW} = 1/ 127.907, \\
m_Z = 91.1876 \textrm{ GeV}, &&\quad
m_t = 165.0 \textrm{ GeV}, 
\eea
which give rise to the values
\be
g = 0.357,\quad 
g^\prime = 0.652,\quad
\phi_0 = 173.458 \textrm{ GeV}, 
\ee
where $\phi_0$ denotes the $T=0$ value of the vev $\phi$.

If not stated differently, we have used the SUSY-breaking parameters
\bea
m_U = m_Q = 500 \textrm{ GeV},&&\quad 
a_t = 100 \textrm{ GeV},\quad \nn \\
M_2 = 2 M_1 = 200 \textrm{ GeV}, &&\quad 
m_E = 500 \textrm{ GeV},
\eea
which enter in the mass matrices that we will define subsequently.

\subsection{Higgs bosons}

For the neutral Higgs bosons we use the notation
\be 
H^0_i = \rme^{iq_i}(\phi_i + S_i + iP_i), \quad 
S = \rme^{iq_s}(\phi_s + S_s + iP_s),
\ee
and $q_1 = q_2 = q/2$. The corresponding mass matrix has
the following form
\bea
M_H^2 = \begin{pmatrix}
M^2_{SS} & M^2_{SP} \\
M^2_{PS} & M^2_{PP}
\end{pmatrix},
\eea
with the matrices $M_{SS}$, $M_{SP}=M_{PS}^\dagger$, and $M_{PP}$
given by the following entries. The CP-even entries at the tree-level read
\bea
M^2_{SS11} &=& m_1^2 + \lambda^2 (\phi_2^2+\phi_s^2) + \frac{\bar g^2}4 (3\phi_1^2- \phi_2^2), \nn \\
M^2_{SS12} &=& -m_{12}^2 \cos(q) + a_\lambda \phi_s \cos(q+q_s)
 		+ \frac12 \phi_1\phi_2(4\lambda^2 - \bar g^2), \nn \\
M^2_{SS13} &=& a_\lambda \phi_2 \cos(q+q_s) + 2\lambda^2\phi_1\phi_s, \nn \\
M^2_{SS22} &=& m_2^2 + \lambda^2 (\phi_1^2+\phi_s^2) + \frac{\bar g^2}4 (3\phi_2^2- \phi_1^2), \nn \\
M^2_{SS23} &=& a_\lambda \phi_1 \cos(q+q_s) + 2\lambda^2\phi_2\phi_s, \nn \\
M^2_{SS33} &=& m_s^2 + \lambda^2 \phi^2.
\eea
The CP-odd entries are
\bea
M^2_{PP11} &=& m_1^2 + \lambda^2 (\phi_2^2+\phi_s^2) + \frac{\bar g^2}4 (\phi_1^2- \phi_2^2), \nn \\
M^2_{PP12} &=& m_{12}^2 \cos(q) - a_\lambda \phi_s \cos(q+q_s), \nn \\
M^2_{PP13} &=& - a_\lambda \phi_2 \cos(q+q_s),  \nn \\
M^2_{PP22} &=& m_2^2 + \lambda^2 (\phi_1^2+\phi_s^2) + \frac{\bar g^2}4 (\phi_2^2- \phi_1^2), \nn \\
M^2_{PP23} &=& - a_\lambda \phi_1 \cos(q+q_s), \nn \\
M^2_{PP33} &=& m_s^2 + \lambda^2 \phi^2.
\eea
Finally the CP-mixed entries yield
\bea
M^2_{SP12} &=& m_{12}^2 \sin(q) - a_\lambda \phi_s \sin(q+q_s), \nn \\
M^2_{SP13} &=& - a_\lambda \phi_2 \sin(q+q_s),  \nn \\
M^2_{SP23} &=& - a_\lambda \phi_1 \sin(q+q_s). \label{H_mixed}
\eea
In the CP-conserving case the submatrix (\ref{H_mixed}) vanishes so that 
CP-even and CP-odd states do not mix.

If the one-loop effective potential is included, we determine the 
masses of the neutral Higgses by the second derivatives of the effective
potential. 

The mass matrix of the charged Higgs bosons in the basis $(H_1^-, \bar H_2^+)$ 
contains the complex entries
\bea
M^2_{H^\pm_{11}} &=& m_1^2 + \lambda^2 \phi -\frac{\bar g^2}4 (\phi_2^2-\phi_1^2)
	+ \frac{g^2}2 \phi_2^2, \nn \\
M^2_{H^\pm_{12}} &=& -\frac12 \phi_1\phi_2 (2\lambda^2 - g^2) 
	+ m^2_{12} \rme^{iq} - a_\lambda \phi_s \rme^{i(q + q_s)}, \nn \\
M^2_{H^\pm_{22}} &=& m_2^2 + \lambda^2 \phi -\frac{\bar g^2}4 (\phi_1^2-\phi_2^2)
	+ \frac{g^2}2 \phi_1^2. 
\eea 

\subsection{Charginos and Neutralinos}

The chargino mass matrix reads
\bea
M_{\tilde\chi^\pm} = \begin{pmatrix}
0 & Y_{\tilde\chi^\pm}^T \\
Y_{\tilde\chi^\pm} & 0 \\
\end{pmatrix}, \quad
Y_{\tilde\chi^\pm} = \begin{pmatrix}
M_2 & g\phi_2 \rme^{-i\frac{q}2} \\
g\phi_1 \rme^{-i\frac{q}2} & -\lambda \phi_s \rme^{i q_s} \\
\end{pmatrix},
\eea
and is diagonalized by the biunitary transformation 
$Y^{\rm diag}_{\tilde\chi^\pm} = U^* Y_{\tilde\chi^\pm} V^\dagger$.

The symmetric mass matrix of the neutralinos in the 
basis $\tilde\chi^0 = (i\tilde B, i \tilde W^3, \tilde H_1^0, \tilde H_2^0, \tilde S)$ yields
\bea
M_{\tilde\chi^0} = 
\begin{pmatrix}
M_1 & .&.&.&. \\
0& M_2 &. &. &. \\
- \frac{g^\prime}{\sqrt{2}} \phi_1 \rme^{-i\frac{q}2} & 
 \frac{g}{\sqrt{2}} \phi_1 \rme^{-i\frac{q}2} & 0 &. &. \\
 \frac{g^\prime}{\sqrt{2}} \phi_2 \rme^{-i\frac{q}2} & 
 - \frac{g}{\sqrt{2}} \phi_2 \rme^{-i\frac{q}2} & \lambda \phi_s \rme^{i q_s}  & 0 &. \\
0 & 0 & \lambda \phi_2 \rme^{i\frac{q}2} & \lambda \phi_1 \rme^{i\frac{q}2} & 0
\end{pmatrix},
\eea
and can be diagonalized using a unitary matrix, 
$M^{\rm diag}_{\tilde\chi^0}=X^T M_{\tilde\chi^0} X$.

\subsection{Gauge bosons}

The mass of the $W$-boson is given by
\be
m^2_W = \frac12 g^2 \phi^2,
\ee
while the photon and the $Z$-boson share the following 
hermitian mass matrix
\be
M^2_{Z\gamma} = \begin{pmatrix}
\frac12 g^2\phi^2 &  -\frac12 gg^\prime \phi^2 \\
 -\frac12 gg^\prime \phi^2 &  \frac12 {g^\prime}^2\phi^2
\end{pmatrix}
\ee
that leads to the $Z$-boson mass $m_Z= ({\bar g}/{\sqrt{2}}) \phi$.

\subsection{Tops and stops}

The top quark has the mass
\be
m^2_t = y_t^2 \phi_2^2,
\ee
and the masses of the stops are given by the following hermitian matrix
\be
M^2_{\tilde t} = \begin{pmatrix}
m_Q^2 + m_t^2 + \frac14(g^2- \frac13 {g^\prime}^2)(\phi_1^2- \phi_2^2) &  
a_t\phi_2 \rme^{iq/2} + y_t \lambda \phi_s\phi_1 \rme^{-i(q_s+q/2)} \\
a_t\phi_2 \rme^{-iq/2} + y_t \lambda \phi_s\phi_1 \rme^{i(q_s+q/2)} & 
m_U^2 + m_t^2 + \frac13 {g^\prime}^2(\phi_1^2- \phi_2^2)
\end{pmatrix}.
\ee

\subsection{Sneutrinos and selectrons}

For the selectrons we have
\be
M^2_{\tilde e} = \begin{pmatrix}
m_E & -\lambda \phi_s e^{iq_s} m_e \, \tan(\beta)\\
-\lambda \phi_s e^{-iq_s} m_e \, \tan(\beta) & 
m_E
\end{pmatrix},
\ee
that is diagonalized via the transformation 
$M^{2}_{\tilde e,\rm diag} = D^\dagger M^2_{\tilde e} D$.

For the electron EDM contributions from the charginos we use a sneutrino of mass
$m_{\tilde \nu} = 1$~TeV. 

\section{One-loop Contributions to the Electron EDM\label{app_edm}}

In this Appendix we briefly discuss the one-loop contributions to
the electron EDM coming from chargino and neutralino exchange.  For the 
sfermions of the first two generations we assume masses of 1 TeV.
The contribution from the charginos is given by~\cite{Ibrahim:1998je}:
\bea
\frac{d_{\rm e-chargino}}{e} = \frac{\alpha_{EM}}{4\pi \sin^2 (\theta_W)} 
\sum_{i=1}^2 \frac{m_{\tilde\chi_i^+}}{m^2_{\tilde \nu_e}} \textrm{ Im}(\Gamma_{ei}) 
\textrm{ A}\bigg(\frac{m^2_{\tilde\chi_i^+}}{m^2_{\tilde \nu_e}}\bigg),
\eea
where 
\be
\Gamma_{ei} = \kappa_e U^*_{i2}V_{i1}, \quad 
\kappa_e = \frac{m_e \, e^{iq/2}}{\sqrt{2} m_W \cos(\beta)}.  
\ee
Here, the matrices $U$ and $V$ diagonalize the chargino mass matrix as defined in the 
previous section and
\be
\textrm{ A}(r) = \frac{1}{2(1-r)^2} \left(3-r + \frac{2}{1-r} \ln (r) \right)
\ee
denotes the loop function.

Analogously, the contribution from the neutralinos 
is~\cite{Ibrahim:1998je,delAguila:1983kc}
\bea
d_{\rm e-neutralino}/e = \frac{\alpha_{EM}}{4\pi \sin^2 (\theta_W)} 
\sum_{k=1}^2\sum_{i=1}^4 \frac{m_{\tilde\chi_i^0}}{m^2_{\tilde e}} \textrm{ Im}(\eta_{eik}) 
\textrm{ B}(m^2_{\tilde\chi_i^0} / m^2_{\tilde e}),
\eea
with 
\bea
\eta_{eik} &=& \left[ 
(\tan (\theta_W) X_{1i} + X_{1i}) D_{1k}^*  
+ \sqrt{2} \kappa_e X_{3i} D_{2k}^*  
\right] \nn \\ 
&& \times \left[ 
\tan (\theta_W) X_{1i}  D_{2k}^*   
+ \frac{\kappa_e}{\sqrt{2}} X_{3i} D_{1k}^*   \right]. 
\eea
In this equation, $X$ and $D$ diagonalize the neutralinos and selectrons, respectively.
The loop function B is defined by
\be
\textrm{ B}(r) = \frac{1}{2(1-r)^2} \left(1+r +\frac{2r}{1-r} \ln (r) \right).
\ee
Following the discussion in Ref.~\cite{Ibrahim:1998je}, it can be
shown that if $q_t$ is the only CP-violating phase in the Higgs
sector, both contributions depend only on the CP-odd combination
$(q_s + q)$. The current experimental bound on the electron EDM
is $d_e < 1.6 \times 10^{-27} \,e \, {\rm cm}$~ \cite{Regan:2002ta}.

\section{Example sets\label{app_ex}}

In this appendix we give some examples of parameters that develop a
strong first-order phase transition. The examples are not chosen
arbitrarily, but they represent specific cases. The first set, shown
in Tables~\ref{example_param}--\ref{example_pt}, accomplishes the
generation of a large BAU thanks to a relatively thin wall. On the other hand
the electron EDM is extremely small, partly because the combination
$q+q_s$ is rather small in the broken phase and partly due to a
coincidental cancellation between the neutralino and the chargino
contributions to the electron EDM. Set number 6 describes a model
with small $\lambda$, taken from Fig.~\ref{fig_al} with
$a_\lambda=177.3$ GeV.  This model generates more than the observed
BAU, but the calculated electron EDM is slightly too big, so that the
sfermions of the first two generations have to be heavier than 1 TeV.
Starting from $q_t~\sim0.1$ instead of 0.3 should yield $\eta_{10}\sim1$
and an electron EDM within the experimental bound.
The remaining sets are randomly chosen. Set number 6 demonstrates that even 
a sizable value of $q_t$ does not lead to a large baryon asymmetry if $\Delta q_s$
is small along the bubble wall. The rather thick  wall induces an
additional suppression of  $\eta_{10}$ in this case.

\begin{table}[p]
\begin{tabular}{|c|| c |c| c| c| c| c | c | c |}
\hline 
 set & \, $m_1$ in GeV \, & \,$m_2$ in GeV \, & \,$m_{12}$ in GeV \, & 
\, $m_s$ in GeV \, & \quad\,$\lambda$ \quad\, &\, $a_\lambda$ in GeV  \,&
\, $t_s^{1/3}$ in GeV \, &\,\,\,\, $q_t$\,\,\, \,\\
\hline 
\hline 
1 & 157.2 & 93.0   & 170.6   & 55.8  & 1.2642   & 268.2   & 98.4   & 2.113 \\
2 & 124.2 & 149.2   & 215.0   & 127.2  & 1.4320   & 254.1   & 152.2   & 5.050\\
3 & 248.9 & 230.8   & 243.3   & 160.5  & -0.8937   & 214.2   & 190.5   & 0.046\\
4 & 397.1 & 251.9   & 375.5   & 342.1  & -1.2991   & 292.0   & 310.5   & 0.282\\
5 & 66.4 & 98.5   & 111.8   & 65.5  & 0.7656   & 95.3   & 102.0   & 6.193\\
6 & 425.1 & 165.7   & 240.1   & 26.8  & 0.5500   & 177.3   & 70.9   & 0.300 \\
\hline 
\end{tabular}
\caption{Parameter examples used for the numerical analysis.\label{example_param}}
\vskip 0.5cm

\begin{tabular}{| c || c | c | c | c | c |}
\hline 
 set &  
\multicolumn{2}{c|}{$\tilde \chi^\pm$} &
\quad $H^\pm$ \quad &
\multicolumn{2}{c|}{$\tilde t_{1/2}$}  \\
\hline 
\hline 
1 & \, 221.77 \, & \,107.40 \, & \,219.56 \, & \,521.43 \, & \,529.30 \, \\
2 & \, 221.64 \, & \,131.22 \, & \,236.41 \, & \,537.16 \, & \,514.91 \, \\
3 & \, 270.45 \, & \,109.31 \, & \,410.84 \, & \,485.74 \, & \,563.03 \, \\
4 & \, 146.21 \, & \,315.37 \, & \,588.45 \, & \,479.29 \, & \,567.75 \, \\
5 & \, 221.06 \, & \,105.94 \, & \,148.97 \, & \,530.64 \, & \,521.67 \, \\
6 & \, 225.24 \, & \,144.62 \, & \,494.87 \, & \,520.50 \, & \,528.89 \, \\
\hline 
\end{tabular}
\caption{The mass spectrum of the charginos, charged Higgses and stops (in GeV).\label{example_m1}}
\vskip 0.5cm

\begin{tabular}{| c || c | c | c | c | c |}
\hline 
 set &  
\multicolumn{5}{|c|}{$S$, $P$} \\
\hline 
\hline 
1 & \, 142.72 \, & \,210.53 \, & \,217.41 \, & \,273.78 \, & \,357.83 \, \\
2 & \, 177.85 \, & \,260.59 \, & \,292.50 \, & \,296.78 \, & \,396.55 \, \\
3 & \, 119.12 \, & \,202.13 \, & \,251.16 \, & \,432.44 \, & \,455.77 \, \\
4 & \, 121.65 \, & \,395.09 \, & \,448.97 \, & \,602.98 \, & \,641.00 \, \\
5 & \, 115.07 \, & \,118.67 \, & \,169.87 \, & \,211.08 \, & \,237.02 \, \\
6 & \, 76.74 \, & \,89.86 \, & \,114.76 \, & \,504.68 \, & \,506.91 \, \\
\hline 
\end{tabular}
\caption{The mass spectrum of the neutral Higgses (in GeV).\label{example_m2}}
\end{table}
\vskip 0.5cm

\begin{table}[t]
\begin{tabular}{| c || c | c | c | c | c |}
\hline 
 set &  
\multicolumn{5}{|c|}{$\tilde \chi^0$} \\
\hline 
\hline 
1 & \, 267.76 \, & \,105.80 \, & \,113.09 \, & \,221.89\, & \, 184.03  \, \\
2 & \, 313.48 \, & \,105.14 \, & \,138.79 \, & \,222.79\, & \, 198.12  \, \\
3 & \, 135.40 \, & \,68.73 \, & \,89.48 \, & \,275.43 \, & \,269.04  \, \\
4 & \, 389.60 \, & \,323.06 \, & \,81.50 \, & \,161.87 \, & \,123.56  \, \\
5 & \, 222.16 \, & \,181.83 \, & \,116.61 \, & \,107.33 \, & \,94.72  \, \\
6 & \, 227.33 \, & \,42.25  \, & \,105.37 \, & \,165.39 \, & \,175.04 \, \\
\hline 
\end{tabular}
\caption{The mass spectrum of the neutralinos (in GeV).\label{example_m3}}
\vskip 0.5cm

\begin{tabular}{|c|| c |c| c| c| c| c | c | c |}
\hline 
 set &\, $\phi$ in GeV\, & \quad\,$\beta$ \quad\, & 
\quad\,$q$ \quad\,& \,$\phi_s$ in GeV \,& \,\quad$q_s$\quad\, \\
\hline 
\hline 
1 & 173.46 & 0.926   & 0.157   & -71.0  & -0.392\\
2 & 173.46 & 0.739   & -0.214   & -81.8  & 0.687\\
3 & 173.46 & 0.808   & 0.015   & -202.9  & -0.036\\
4 & 173.46 & 0.915   & 0.074   & -202.6  & -0.256\\
5 & 173.46 & 0.714   & -0.024   & -112.6  & 0.051\\
6 & 173.46 & 1.108   & 0.029   & -251.9  & -0.067 \\
\hline 
\end{tabular}
\caption{The vevs in the broken phase at temperature $T=0$.\label{example_x_0}}
\vskip 0.5cm

\begin{tabular}{|c|| c |c| c| c| c| c | c | c |}
\hline 
 set &\, $\phi$ in GeV\, & \quad\,$\beta$ \quad\, & 
\quad\,$q$ \quad\,& \,$\phi_s$ in GeV \,& \,\quad$q_s$\quad\, \\
\hline 
\hline 
1 & 165.4 & 0.900   & 0.175   & -71.4  & -0.437\\
2 & 170.0 & 0.726   & -0.223   & -83.2  & 0.708\\
3 & 150.3 & 0.803   & 0.016   & -214.9  & -0.038\\
4 & 164.5 & 0.911   & 0.076   & -207.4  & -0.259\\
5 & 151.6 & 0.687   & -0.029   & -120.2  & 0.058\\
6 & 141.9 & 1.102   & 0.041   & -261.5  & -0.092\\
\hline 
\end{tabular}
\caption{The vevs in the broken phase at temperature $T=T_c$.\label{example_x_brok}}
\end{table}
\vskip 0.5cm

\begin{table}[t]
\begin{tabular}{|c|| c |c| c| c| c| c | c | c |}
\hline 
 set &\, $\phi$ in GeV\, & \quad\,$\beta$ \quad\, & 
\quad\,$q$ \quad\,& \,$\phi_s$ in GeV \,& \,\quad$q_s$\quad\, \\
\hline 
\hline 
1 &  0.0 & -   & -   & -281.4  & -2.113\\
2 &  0.0 & -   & -   & -241.6  & 1.233\\
3 &  0.0 & -   & -   & -283.3  & -0.046\\
4 &  0.0 & -   & -   & -259.9  & -0.282\\
5 &  0.0 & -   & -   & -266.0  & 0.090\\
6 &  0.0 & -   & -   & -471.5  & -0.300 \\
\hline 
\end{tabular}
\caption{The vevs in the symmetric phase at temperature $T=T_c$.\label{example_x_symm}}
\vskip 0.5cm

\begin{tabular}{|c|| c |c| c| c| c| c | c | c |}
\hline 
 set & \,$T_c$ in GeV \,& \,$d_e$ in $10^{-27} \,e \,cm \,$&
	\ $l_w$ in $T_c^{-1}$\ & \quad\, $\eta_{10}$ \quad\, \\
\hline 
\hline 
1 & 113.5 & 0.002 & 2.43 & 29.443\\
2 & 99.1 & 0.796 & 6.38 & -3.214\\
3 & 109.1 & 0.499 & 7.82 & 0.014\\
4 & 78.6 & 5.894 & 34.01 & 0.005\\
5 & 115.8 & -0.893 & 3.05 & -0.398\\
6 & 105.6 & -2.054 & 2.39 & 2.717 \\
\hline 
\end{tabular}
\caption{The parameters of the phase transition and the generated BAU.\label{example_pt}}
\end{table}


\clearpage

\begin{thebibliography}{99}
\bibliographystyle{unsrt}

\bibitem{Kuzmin:1985mm}
  V.~A.~Kuzmin, V.~A.~Rubakov and M.~E.~Shaposhnikov,
  ``On the anomalous electroweak baryon number nonconservation in the early
  universe'',
  Phys.\ Lett.\ B {\bf 155} (1985) 36.

\bibitem{Cohen:1994ss}
  A.~G.~Cohen, D.~B.~Kaplan and A.~E.~Nelson,
  ``Diffusion enhances spontaneous electroweak baryogenesis'',
  Phys.\ Lett.\ B {\bf 336} (1994) 41
  [hep-ph/9406345].

\bibitem{Joyce:1994fu}
  M.~Joyce, T.~Prokopec and N.~Turok,
  ``Electroweak baryogenesis from a classical force'',
  Phys.\ Rev.\ Lett.\  {\bf 75} (1995) 1695
  [Erratum ibid.\  {\bf 75} (1995) 3375]
  [hep-ph/9408339].

  M.~Joyce, T.~Prokopec and N.~Turok,
  ``Nonlocal electroweak baryogenesis. Part 2: the classical regime'',
  Phys.\ Rev.\ D {\bf 53} (1996) 2958
  [hep-ph/9410282].

\bibitem{JCK}
J.M.~Cline, M.~Joyce and K.~Kainulainen, 
``Supersymmetric electroweak baryogenesis in the WKB approximation'',
Phys. Lett. {\bf B417} (1998) 79,
Erratum ibid. {\bf B448} (1999) 321 [hep-ph/9708393]. 

J.~M.~Cline, M.~Joyce and K.~Kainulainen,
  ``Supersymmetric electroweak baryogenesis'',
  JHEP {\bf 0007} (2000) 018
  [hep-ph/0006119].

  J.~M.~Cline, M.~Joyce and K.~Kainulainen,
  ``Supersymmetric electroweak baryogenesis. (Erratum)'',
  hep-ph/0110031.

\bibitem{Kainulainen:2001cn}
  K.~Kainulainen, T.~Prokopec, M.~G.~Schmidt and S.~Weinstock,
  ``First principle derivation of semiclassical force for electroweak
  baryogenesis'',
  JHEP {\bf 0106} (2001) 031
  [hep-ph/0105295].

  T.~Prokopec, M.~G.~Schmidt and S.~Weinstock,
  ``Transport equations for chiral fermions to order h-bar and electroweak
  baryogenesis'',
  Ann.~Phys.\  {\bf 314} (2004) 208
  [hep-ph/0312110].

  T.~Prokopec, M.~G.~Schmidt and S.~Weinstock,
  ``Transport equations for chiral fermions to order h-bar and electroweak
  baryogenesis. II'',
  Ann.~Phys.\  {\bf 314} (2004) 267
  [hep-ph/0406140].

\bibitem{kpss}
  T.~Konstandin, T.~Prokopec, M.~G.~Schmidt and M.~Seco,
  ``MSSM electroweak baryogenesis and flavour mixing in transport  equations'',
  Nucl.\ Phys.\ B {\bf 738} (2006) 1
  [hep-ph/0505103].

\bibitem{cmqsw2}
  M.~Carena, M.~Quiros, M.~Seco and C.~E.~M.~Wagner,
  ``Improved results in supersymmetric electroweak baryogenesis'',
  Nucl.\ Phys.\ B {\bf 650} (2003) 24
  [hep-ph/0208043].

\bibitem{cmqsw}
  M.~Carena, J.~M.~Moreno, M.~Quiros, M.~Seco and C.~E.~M.~Wagner,
  ``Supersymmetric CP-violating currents and electroweak baryogenesis'',
  Nucl.\ Phys.\ B {\bf 599} (2001) 158
  [hep-ph/0011055].


\bibitem{kps}
  T.~Konstandin, T.~Prokopec and M.~G.~Schmidt,
  ``Kinetic description of fermion flavor mixing and CP-violating sources  for
  baryogenesis'',
  Nucl.\ Phys.\ B {\bf 716} (2005) 373
  [hep-ph/0410135].

\bibitem{Joyce:1999fw}
  M.~Joyce, K.~Kainulainen and T.~Prokopec,
  ``The semiclassical propagator in field theory'',
  Phys.\ Lett.\ B {\bf 468} (1999) 128
  [hep-ph/9906411].

\bibitem{FH06}
L.~Fromme and S.J.~Huber,
``Top transport in electroweak baryogenesis'',
hep-ph/0604159.


\bibitem{Moreno:1998bq}
  J.~M.~Moreno, M.~Quiros and M.~Seco,
  ``Bubbles in the supersymmetric standard model'',
  Nucl.\ Phys.\ B {\bf 526}, 489 (1998)
  [hep-ph/9801272].

\bibitem{PT98}
C.~Panagiotakopoulos and K.~Tamvakis,
``Stabilized NMSSM without domain walls'',
Phys.~Lett.~{\bf B446} (1999)  224
  [hep-ph/9809475].

\bibitem{PT99}
C.~Panagiotakopoulos and K.~Tamvakis,
``New minimal extension of MSSM'',
Phys.~Lett.~{\bf B469} (1999) 145
[hep-ph/9908351].

\bibitem{Panagiotakopoulos:2000wp}
  C.~Panagiotakopoulos and A.~Pilaftsis,
  ``Higgs scalars in the minimal non-minimal supersymmetric standard model'',
  Phys.\ Rev.\ D {\bf 63} (2001) 055003
  [hep-ph/0008268].

\bibitem{DHMT00}
A.~Dedes, C.~Hugonie, S.~Moretti and K.~Tamvakis, 
``Phenomenology of a new minimal supersymmetric extension of the standard model'',
Phys.~Rev.~{\bf D63} (2001) 055009
[hep-ph/0009125].

\bibitem{ASW95}
S.A.~Abel, S.~Sarkar and P.L.~White,
``On the cosmological domain wall problem for the minimally extended 
supersymmetric standard model'',
Nucl.~Phys.~{\bf B454} (1995) 663
[hep-ph/9506359].


\bibitem{Menon:2004wv}
  A.~Menon, D.~E.~Morrissey and C.~E.~M.~Wagner,
  ``Electroweak baryogenesis and dark matter in the nMSSM'',
  Phys.\ Rev.\ D {\bf 70} (2004) 035005
  [hep-ph/0404184].

\bibitem{Chang:2002ex}
D.~Chang, W.~F.~Chang and W.~Y.~Keung,
``New constraint from electric dipole moments on chargino baryogenesis in
MSSM'',
Phys.\ Rev.\ D {\bf 66} (2002) 116008
[hep-ph/0205084].

\bibitem{Pilaftsis:2002fe}
A.~Pilaftsis,
``Higgs-mediated electric dipole moments in the MSSM: an application to
baryogenesis and Higgs searches'',
Nucl.\ Phys.\ B {\bf 644}, (2002) 263
[hep-ph/0207277].

\bibitem{P92}
M.~Pietroni,
``The electroweak phase transition in a nonminimal supersymmetric model'',
Nucl.~Phys.~{\bf B402} (1993) 27
[hep-ph/9207227].

\bibitem{DFM96}
A.T.~Davies, C.D.~Froggatt and R.G.~Moorhouse,
``Electroweak baryogenesis in the next-to-minimal supersymmetric model'',
Phys.~Lett.~{\bf B372} (1996) 88
[hep-ph/9603388].

\bibitem{HS98}
S.J.~Huber and M.G.~Schmidt,
``SUSY variants of the electroweak phase transition'',
Eur.~Phys.~J.~{\bf C10} (1999) 473
[hep-ph/9809506].

\bibitem{BHKRV00}
M.~Bastero-Gil, C.~Hugonie, S.F.~King, D.P.~Roy and S.~Vempati, 
``Does LEP prefer the NMSSM?'',
Phys.~Lett.~{\bf B489} (2000) 359
[hep-ph/0006198].


\bibitem{KLT04}
J.~Kang, P.~Langacker and T.~Li,
``Electroweak baryogenesis in a supersymmetric U(1)-prime model'',
Phys.~Rev.~Lett.~{\bf 94} (2005) 061801
[hep-ph/0402086].

\bibitem{Huber:2000mg}
  S.~J.~Huber and M.~G.~Schmidt,
  ``Electroweak baryogenesis: concrete in a SUSY model with a gauge  singlet'',
  Nucl.\ Phys.\ B {\bf 606} (2001) 183
  [hep-ph/0003122].

  S.~J.~Huber, P.~John, M.~Laine and M.~G.~Schmidt,
  ``CP violating bubble wall profiles'',
  Phys.\ Lett.\ B {\bf 475} (2000) 104
  [hep-ph/9912278].

\bibitem{Funakubo:2005pu}
  K.~Funakubo, S.~Tao and F.~Toyoda,
  ``Phase transitions in the NMSSM'',
  Prog.\ Theor.\ Phys.\  {\bf 114} (2005) 369
  [hep-ph/0501052].

\bibitem{Bodeker:1996pc}
D.~Bodeker, P.~John, M.~Laine and M.~G.~Schmidt,
``The 2-loop MSSM finite temperature effective potential with stop
condensation'',
Nucl.\ Phys.\ B {\bf 497}, (1997) 387
[hep-ph/9612364].

\bibitem{Laine:2000rm}
M.~Laine and K.~Rummukainen,
``Two Higgs doublet dynamics at the electroweak phase transition: a
non-perturbative study,''
Nucl.\ Phys.\ B {\bf 597},  (2001) 23
[hep-lat/0009025].

\bibitem{LEP}
LEP Collaborations, ALEPH Collaboration, DELPHI Collaboration, L3
Collaboration, OPAL Collaboration and Line Shape Sub-Group of the LEP
Electroweak Working Group, 
``Combination procedure for the precise
determination of Z boson parameters from results of the LEP
experiments'', hep-ex/0101027.

\bibitem{HKLM03}
R.~Harnik, G.D.~Kribs, D.T.~Larson and H.Murayama,
``The minimal supersymmetric fat Higgs model'',
Phys.~Rev.~{\bf D70} (2004) 015002
[hep-ph/0311349].

\bibitem{DT05}
A.~Delgado and T.M.P. Tait, 
``A fat Higgs with a fat top'',
JHEP {\bf 0507} (2005) 023
[hep-ph/0504224].

\bibitem{Cirigliano:2006dg}
V.~Cirigliano, S.~Profumo and M.~J.~Ramsey-Musolf,
``Baryogenesis, electric dipole moments and dark matter in the MSSM'',
hep-ph/0603246.

\bibitem{Huet:1995sh}
  P.~Huet and A.~E.~Nelson,
  ``Electroweak baryogenesis in supersymmetric models'',
  Phys.\ Rev.\ D {\bf 53},  (1996) 4578
  [hep-ph/9506477].

\bibitem{Carena:1997gx}
  M.~Carena, M.~Quiros, A.~Riotto, I.~Vilja and C.~E.~M.~Wagner,
  ``Electroweak baryogenesis and low energy supersymmetry'',
  Nucl.\ Phys.\ B {\bf 503} (1997) 387
  [hep-ph/9702409].






\bibitem{Lee:2004we}
  C.~Lee, V.~Cirigliano and M.~J.~Ramsey-Musolf,
  ``Resonant relaxation in electroweak baryogenesis'',
  Phys.\ Rev.\ D {\bf 71} (2005) 075010
  [hep-ph/0412354].

\bibitem{Cirigliano:2006wh}
  V.~Cirigliano, M.~J.~Ramsey-Musolf, S.~Tulin and C.~Lee,
  ``Yukawa and tri-scalar processes in electroweak baryogenesis'',
  hep-ph/0603058.






\bibitem{Regan:2002ta}
B.~C.~Regan, E.~D.~Commins, C.~J.~Schmidt and D.~DeMille,
``New limit on the electron electric dipole moment'',
Phys.\ Rev.\ Lett.\  {\bf 88} (2002) 071805.

\bibitem{Baker:2006ts}
  C.~A.~Baker et al.,
  ``An improved experimental limit on the electric dipole moment of the
  neutron'', hep-ex/0602020.

\bibitem{Pospelov:2005pr}
  M.~Pospelov and A.~Ritz,
  ``Electric dipole moments as probes of new physics'',
  Ann.~Phys.\  {\bf 318} (2005) 119
  [hep-ph/0504231].

\bibitem{Barr:1990vd}
  S.~M.~Barr and A.~Zee,
  ``Electric Dipole Moment Of The Electron And Of The Neutron'',
  Phys.\ Rev.\ Lett.\  {\bf 65} (1990) 21
  [Erratum ibid.\  {\bf 65} (1990) 2920].


\bibitem{Chang:1998uc}
  D.~Chang, W.~Y.~Keung and A.~Pilaftsis,
  ``New two-loop contribution to electric dipole moment in supersymmetric
  theories'',
  Phys.\ Rev.\ Lett.\  {\bf 82} (1999) 900
  [Erratum ibid.\  {\bf 83} (1999) 3972]
  [hep-ph/9811202].

\bibitem{MooreProkopec:1995}
  G.~D.~Moore and T.~Prokopec,
  ``How fast can the wall move? A study of the electroweak phase transition
  dynamics'',
  Phys.\ Rev.\ D {\bf 52} (1995) 7182
  [hep-ph/9506475].

  G.~D.~Moore and T.~Prokopec,
  ``Bubble wall velocity in a first order electroweak phase transition'',
  Phys.\ Rev.\ Lett.\  {\bf 75} (1995) 777
  [hep-ph/9503296].

\bibitem{Konstandin:2006nd}
  T.~Konstandin and S.~J.~Huber,
  ``Numerical approach to multi dimensional phase transitions'', 
  JCAP {\bf 0606} (2006) 021
  [hep-ph/0603081].

\bibitem{GSW04}
C.~Grojean, G.~Servant and J.D.~Wells,
``First-order electroweak phase transition in the standard model with a low cutoff'',
Phys.~Rev.~{\bf D71} (2005) 036001
[hep-ph/0407019].

\bibitem{BFHS04}
D.~Bodeker, L.~Fromme, S.J.~Huber and M.~Seniuch,
``The baryon asymmetry in the standard model with a low cut-off'',
JHEP {\bf 0502} (2005) 026
[hep-ph/0412366].

\bibitem{Shaposhnikov:1987pf}
  M.~E.~Shaposhnikov,
  ``Structure of the high temperature gauge ground state and electroweak
  production of the baryon asymmetry'',
  Nucl.\ Phys.\ B {\bf 299},  (1988) 797.

\bibitem{eta}
D.N.~Spergel et al., astro-ph/0603449.

\bibitem{Kainulainen:2002th}
 K.~Kainulainen, T.~Prokopec, M.~G.~Schmidt and S.~Weinstock,
 ``Semiclassical force for electroweak baryogenesis: three-dimensional
 derivation'',
 Phys.\ Rev.\ D {\bf 66} (2002) 043502
 [hep-ph/0202177].

\bibitem{Kang:2004pp}
  J.~Kang, P.~Langacker, T.~j.~Li and T.~Liu,
  ``Electroweak baryogenesis in a supersymmetric U(1)' model''.
  Phys.\ Rev.\ Lett.\  {\bf 94} (2005) 061801
  [hep-ph/0402086].



\bibitem{Ibrahim:1998je}
  T.~Ibrahim and P.~Nath,
  ``The neutron and the lepton EDMs in MSSM, large CP violating phases, and
  the cancellation mechanism'',
  Phys.\ Rev.\ {\bf D58} (1998) 111301
  [Erratum ibid.\ D {\bf 60} (1999) 099902]
  [hep-ph/9807501].



\bibitem{delAguila:1983kc}
F.~del Aguila, M.~B.~Gavela, J.~A.~Grifols and A.~Mendez,
``Specifically supersymmetric contribution to electric dipole moments'',
Phys.\ Lett.\ B {\bf 126} (1983) 71
[Erratum ibid.\ B {\bf 129} (1983) 473].



\end{thebibliography}
\end{document}